\documentstyle[12pt,aaspp4,epsfig]{article}

\begin{document}

\def \cmm  {cm$^{-2}$}
\def \cmmm {cm$^{-3}$}
\def \kms  {km~s$^{-1}$}
\def \lyaf {Ly$\alpha$ forest}
\def \Lya  {Ly$\alpha$}
\def \lya  {Ly$\alpha$}
\def \Lyb  {Ly$\beta$}
\def \lyb  {Ly$\beta$}
\def \Lyg  {Ly$\gamma$}
\def \lyg  {Ly$\gamma$}
\def \ly5  {Ly-5}
\def \ly6  {Ly-6}
\def \ly7  {Ly-7}
\def \nhi  {$N_{HI}$}
\def \ndi  {$N_{DI}$}
\def \lnhi  {$\log N_{HI}$}
\def \lndi  {$\log N_{DI}$}
\def \nhe  {$N_{HeII}$}
\def \object {HS~0105+1619}
\def \qfirst {PKS~1937--1009}
\def \qsecond {Q1009+2956}
\def \qthird {Q0130--4021}
\def \qfourth {HS~0105+1619}
\def \qhst {PG~1718+4807}

\def \ETA {$\eta $}
\def \het {$^3$He}
\def \hef {$^4$He}
\def \lisv {$^7$Li}
\def \yp      {Y$_p$}

\def \btemp {$b_{temp}$}
\def \bturb {$b_{turb}$}
\def \ly    {Lyman}
\def \mf    {$\times 10^{-5}$}

\def \ob {$\Omega _b h^2$}

\title{The Deuterium to Hydrogen Abundance Ratio Towards a Fourth QSO:
HS~0105+1619}

\author{
         John M. O'Meara\altaffilmark{1,2},
         David Tytler\altaffilmark{1,3}, \\
         David Kirkman\altaffilmark{4},
         Nao Suzuki\altaffilmark{5},
         Jason X. Prochaska\altaffilmark{1,6},
         Dan Lubin\altaffilmark{7}
         \& Arthur M. Wolfe\altaffilmark{8}
\\
Center for Astrophysics and Space Sciences;\\
University of California, San Diego; \\
MS 0424; La Jolla; CA 92093-0424\\}

\altaffiltext{1} {Visiting Astronomer, W.M. Keck Observatory which
is a joint facility of the University of California, the California
Institute of Technology and NASA.}
\altaffiltext{2} {E-mail: jomeara@ucsd.edu}
\altaffiltext{3} {E-mail: tytler@ucsd.edu}
\altaffiltext{4} {E-mail: david@mamacass.ucsd.edu}
\altaffiltext{5} {E-mail: suzuki@ucsd.edu}
\altaffiltext{6} {E-mail: xavier@ociw.edu, Observatories of the
Carnegie Institute of Washington}
\altaffiltext{7} {E-mail: dlubin@ucsd.edu}
\altaffiltext{8} {E-mail: awolfe@ucsd.edu}

\abstract
We report the measurement of the primordial D/H abundance ratio
towards QSO \object.  The column density of the neutral hydrogen in
the $z \simeq 2.536$ Lyman limit system is high, \lnhi\ $= 19.422
\pm 0.009$ \cmm , allowing for the deuterium to be seen in 5 Lyman series
transitions.  
The measured value of the D/H ratio towards QSO
\object\ is found to be D/H$ = 2.54 \pm 0.23 \times 10^{-5}$.  
The metallicity of the system showing D/H is found to be $\simeq 0.01$
solar, indicating that the measured D/H is the 
primordial D/H within the measurement errors.  
The gas which shows D/H is neutral, 
unlike previous D/H systems which were more highly
ionized.  Thus, the determination of the D/H ratio becomes more secure
since we are measuring it in different astrophysical environments, but
the error is larger because we now see more dispersion between
measurements.
Combined with prior
measurements of D/H, the best D/H ratio is now
D/H$ = 3.0 \pm 0.4 \times 10^{-5}$, which is 10\% lower than the
previous value.  
The new values for the baryon-to-photon ratio,
and baryonic matter density derived from D/H are $\eta =
5.6 \pm 0.5 \times 10^{-10} $ and \ob $=0.0205 \pm 0.0018$ respectively.


\section{Introduction}

The standard theory of big bang nucleosynthesis (SBBN) predicts the
abundances of the light nuclei H, D, \het , \hef , and \lisv\ as a
function of the cosmological baryon to photon ratio, $\eta =
{n_{b}}/{n_{\gamma}}$ (Kolb \& Turner 1990; Walker et al. 1991;
Schramm \& Turner 1998;
Nollett \& Burles 2000; Olive, Steigman \& Walker 2000).  
A measurement of the ratio of any two
primordial abundances gives $\eta$, and hence the baryon density,
while a second ratio tests the theory.  However, it is extremely
difficult to measure primordial abundances, because in most places
gas ejected from stars has changed the abundances.

Adams (1976) suggested that it might be possible to measure the
primordial D/H ratio in absorption line systems towards QSOs.
Although gas
which has been inside a star will have lost all of its deuterium, in QSO
absorption line systems having typical metal abundances of 0.001 to 0.01 of
the solar value, about 0.1 -- 1\% of the deuterium will have been lost.

The advent of the HIRES spectrograph (Vogt 1994) on the W.M. 
Keck-I telescope gave the high signal-to-noise 
ratio (SNR) and spectral resolution 
needed to reveal deuterium (Tytler et al. 2000) in high
redshift absorption systems.  We have previously measured D/H in two QSOs
(Tytler, Fan \& Burles 1996; 
Tytler \&\ Burles 1997; Burles \&\ Tytler 1998a; Burles \&\
Tytler 1998b), and placed a strong upper limit on D/H in a third
(Kirkman et al. 1999).  

Other QSOs give less useful constraints on D/H, because their
absorption systems are more complex, or existing spectra are
inadequate.  
The Lyman limit system (LLS) at $z_{abs}=0.701$ towards QSO \qhst\ might allow
ten times larger D/H, or it may give no useful constraints (Webb et
al. 1997; Levshakov, Kegel, \& Takahara 1998; Tytler et al. 1999).
Molaro et al. (1999) claimed another QSO absorption system
showed low D/H, but they and Levshakov et al. (2000) note that since
only the \lya\ line has been observed, the hydrogen velocity structure
and the H~I column density
 are poorly known, and the deuterium feature can be fit using hydrogen alone.

In this paper, we present a fourth QSO, \object , which gives strong
constraints on the primordial D/H ratio.

\section{Observations and data reduction}

We report the detection of deuterium in the QSO HS~0105+1619
(emission line redshift 2.64, V=16.9, B1950 RA 1h 5m 26.97s,
DEC +16d 19m 50.1s; J2000 RA 1h 8m 6.4s, DEC +16d 35m 50.0s), which
was discovered by Hagen, Engels \& Reimers (1999), who very kindly gave us
a finding chart and a low resolution spectrum prior to publication.

We present high quality, high SNR spectra of \object\
in both low and high resolution.  The low resolution spectra were
obtained using the Kast double spectrograph on the Shane 3 meter
telescope at Lick observatory.  The high resolution spectra were
obtained using the HIRES spectrograph on the Keck-I telescope.  The
observations are summarized in Table \ref{obstab}.
All of the high resolution observations were taken using the
HIRES image rotator to align the direction of
atmospheric dispersion along the slit,  and were
taken with the C5 decker, which provides an entrance aperture to the
spectrograph with dimensions $1.15^{''} \times 7.5^{''}$ .  The
spectra were sampled in 2.1~\kms\ pixels with the Tektronix 2048x2048
CCD.

\begin{figure}
\centerline{\psfig{file=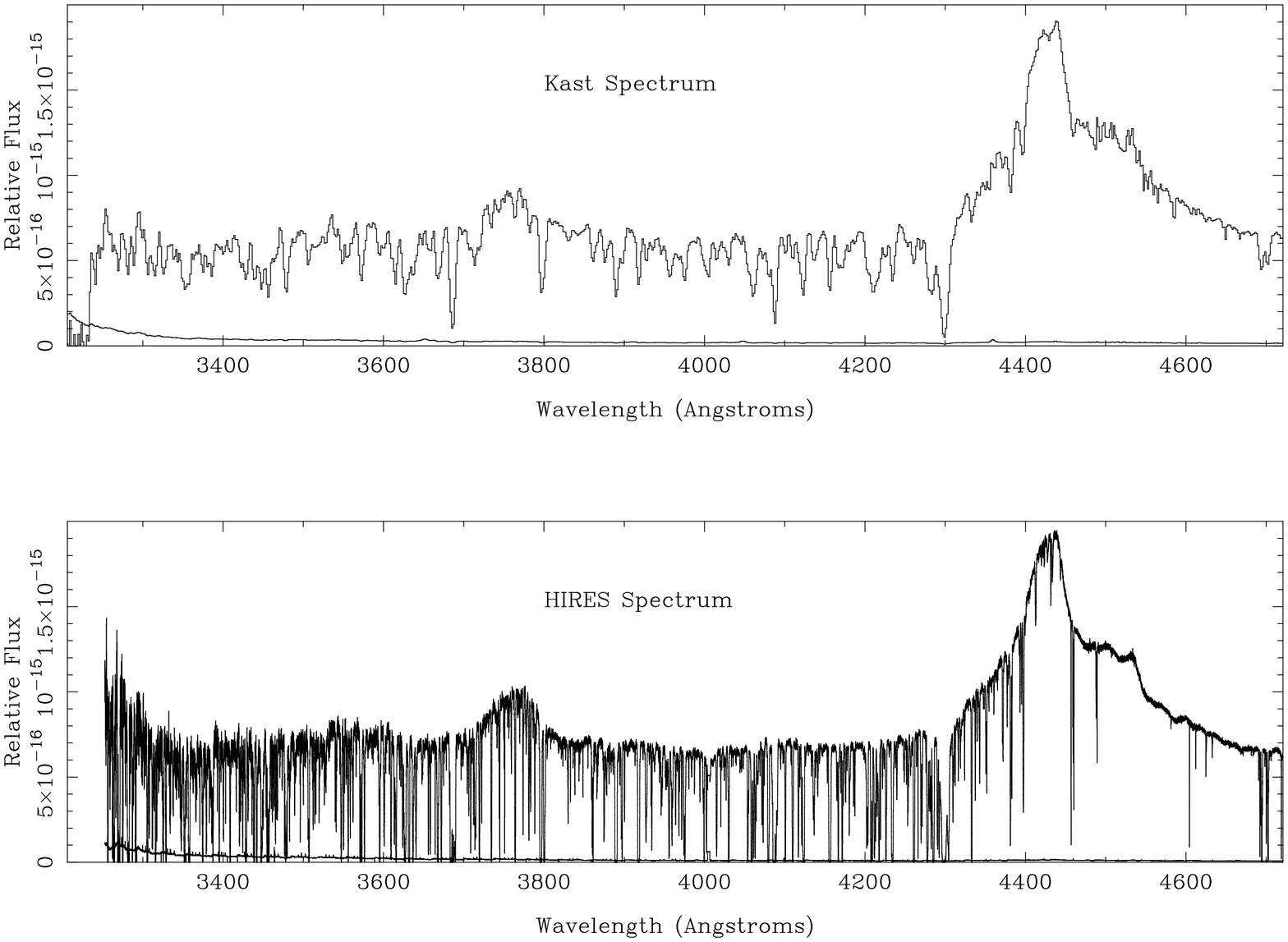, height=6.5in, angle=270}}
\caption{Spectrum of \object .  The upper panel shows the low
resolution flux calibrated spectrum obtained with the Kast spectrograph.
The lower panel shows the flux calibrated HIRES spectrum.  The flux
calibration was noisy at wavelengths less than 3800 \AA\
and was not applied to the Lyman limit, which is not
shown above for the HIRES spectrum.}
\label{lick_keck}
\end{figure}

The HIRES spectra were flat-fielded, optimally extracted, and
wavelength calibrated using Tom Barlow's set of echelle extraction
MAKEE programs.  The spectra were then co-added to produce a
single spectrum.  This spectrum was then flux calibrated using the low
resolution Kast spectrum.  Because the SNR decreases in
both spectra at lower wavelengths, we use the flux calibrated data for
wavelengths greater than 3800 \AA\ and the co-added spectrum without
flux calibration below 3800 \AA . 
The details regarding the co-adding
and fluxing are given in Suzuki \& Tytler (2000).  
The wavelength scale has an external zero point error of approximately
$\pm 10$ \kms , which is the shift between spectra taken at 
different times, which we corrected.
The internal error in the wavelength scale is a least 0.09 \kms , from the
arc line fits, and may be approximately 1 -- 2 \kms , which is the size 
of the error which we have noted in our analysis of other similar spectra
(Levshakov, Tytler \& Burles 2000).
The final spectrum
has 2.1 \kms\ wide pixels, and a SNR of approximately 80 and 10 at the
\lya\ and Lyman limit of the D/H system, respectively. 
In Figure \ref{lick_keck}, we show the flux calibrated
 Kast and HIRES spectra.

\section{General Properties of the $z \simeq 2.536 $ Lyman Limit System Towards \object}
As can be seen in Figure \ref{lick_keck}, the low resolution Kast
spectrum of \object\ shows a steep Lyman
break at a wavelength of approximately 3230 \AA, which is caused by
a Lyman limit system at a redshift of 
$z \simeq 2.54$. At this redshift we see a strong
\lya\ absorption line,  near 4300 \AA. 
Our analysis of the high resolution HIRES spectrum indicates that
there is a Lyman limit system at a redshift of $z \simeq
2.536$, which shows deuterium and numerous metal absorption lines.
We discuss the general features of the absorption below, and the best
fits to the data in the following section. 

\subsection{Hydrogen Absorption}
\begin{figure}
\centerline{\psfig{file=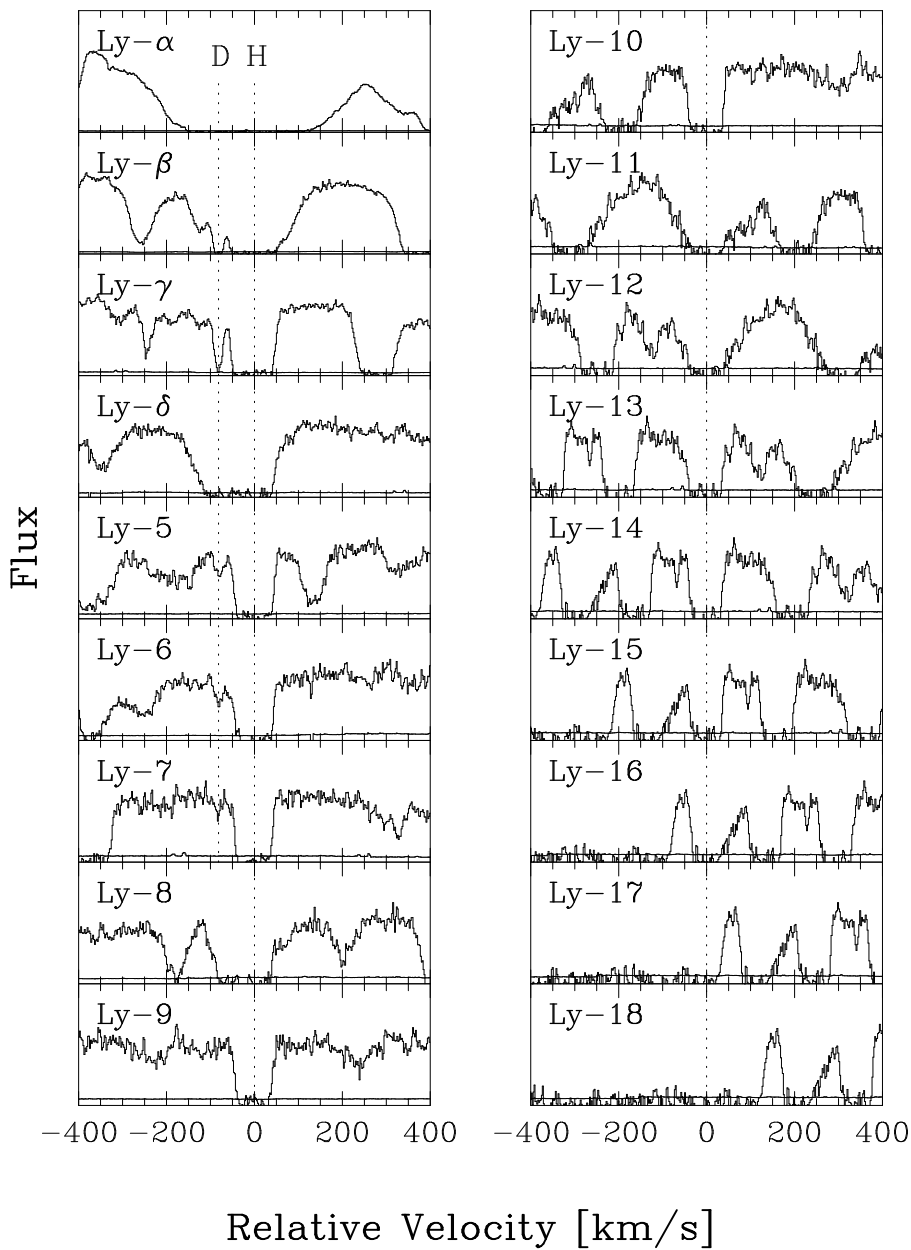, height=6.5in}}
\caption{Lyman series absorption in the $z \simeq 2.536$ Lyman limit
	system towards \object . The velocities shown are relative to
	the H~I redshift of $z=2.535998$.  The vertical scales are
linear flux, from zero, and the lower traces are the $1\sigma$ error.
The D~I absorption is seen at $-82$ \kms\ in \lyb , \lyg , Ly-5, Ly-6
and Ly-7.}
\label{lyseries}
\end{figure}

We observe hydrogen in all Lyman series transitions through
Lyman-17, where the spectrum abruptly ends at 3233 \AA\ due to line
blending near the
Lyman limit.  The observed Lyman transitions are shown in Figure
\ref{lyseries}.  

The column density of the hydrogen is high, as Figure
\ref{lyseries} indicates.  All of the Lyman series transitions are
saturated through to the Lyman limit, indicating that the column
density is at least \lnhi\ $\ge 17.8$ \cmm .  The \lya\ line 
has zero flux across about 200 \kms\, indicating that the column density of the
system is approaching the levels found in damped \lya\
systems.

Inspection of Figure \ref{lyseries} also indicates that the absorption
system is very simple.  The Ly-5, Ly-6, Ly-7, Ly-9, Ly-10, Ly-14, and
Ly-15 transitions all appear symmetric and 
un-blended, allowing us to describe the absorber by a single component.

\subsection{Deuterium Absorption}
Since the column density of the hydrogen responsible for the Lyman
limit appears high, we expect the associated deuterium absorption to be
strong.  
For the first time, we observe absorption at the
predicted position of deuterium, a velocity of
$v = -81.64$ \kms\ in the frame of the H~I,
in 5 Lyman series transitions: \lyb,
\lyg, Ly-5, Ly-6, and Ly-7.  Deuterium \lya\ was not observed since it
is subsumed by the absorption of the hydrogen \lya, and deuterium Ly-4 
was not observed due to intervening \lya\ forest absorption.

The observed absorption is narrow, and appears symmetric and
free of strong contamination, suggesting that like the hydrogen, the
absorption is simple and can be modeled with a single component.

In all transitions where absorption is observed, the velocity of the
absorption appears centered about the same value, $v \simeq -82$ \kms\
, strongly
indicating that the features are all Lyman transitions of the same absorber,
and that the absorber is consistent with being deuterium.  
In a later section, we
give a more rigorous discussion of why we believe the absorption is
indeed deuterium, and not hydrogen or other ions.

\subsection{Metal Line Absorption}
\begin{figure}
\centerline{\psfig{file=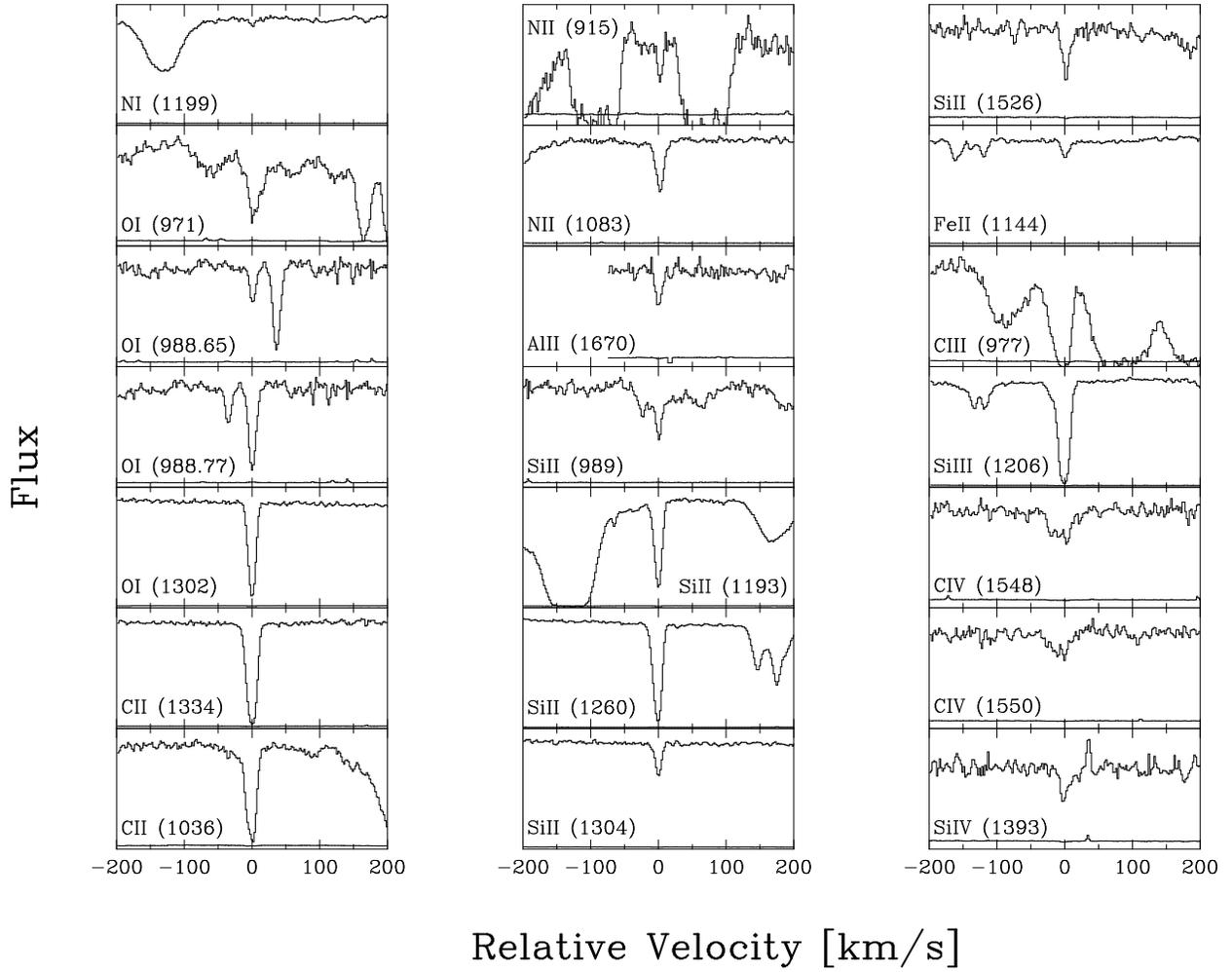, height=6.5in, angle=270}}
\caption{Observed metal line absorption associated with the $z \simeq 2.536$
Lyman limit system.  The metals are grouped according to ionization
state and are organized by atomic mass.  The low ionization lines have
simple, narrow profiles centered near 0 \kms .}
\label{metalfig}
\end{figure}

The $z \simeq 2.536$ Lyman limit system shows a variety of 
metal ions, as seen in Figure \ref{metalfig}.  Unlike Lyman
Limit systems with column densities in the range of \lnhi = 16.5--18 
\cmm\ which show absorption predominantly 
in the higher ionization states, 
this system shows metal line absorption in neutral, low, and high ionization
states.  Since the column density of the hydrogen appears high, we
expect the neutral and low ionization metal lines to trace 
the H I, and with it
the deuterium, in analogy with damped \lya\ systems.

All of the low ionization metal lines are extremely narrow, and
appear to be well described by a single component.
Moreover, the low ions are all centered at $v \simeq 0$ \kms ,
implying that they arise in the same gas as the H~I and D~I.

\section{Best Parameters for the $z \simeq 2.536$ Lyman Limit
System}

We now give the parameters which describe the absorption in the $z
\simeq 2.536 $ Lyman Limit system towards \object .  For all
measurements, a continuum was fit to the region under
consideration to produce a unit normalized spectrum.  The observed
absorption features were then fit using the VPFIT Voigt profile
line-fitting routine (Webb, 1987; kindly provided by Carswell) and
re-verified using in-house routines.  
For each absorber, we obtain an estimate of the column
density, $N$, the redshift, $z$, and the velocity width, $b$, 
along with their
respective $1\sigma$ errors.  The results of this analysis are found
in Table \ref{dhlinetab}.

\subsection{The Hydrogen}
The parameters describing the hydrogen absorption responsible for the Lyman
limit are obtained from various complementary parts of the spectrum,
and are given by \lnhi $= 19.422 \pm 0.009$ \cmm , 
$b=13.99 \pm 0.20$ \kms , and
$z=2.535998 \pm 0.000007$.

The deuterium  and low ionization metal absorption lines 
give a strong indication that
the H~I can be modeled by a single component, whose redshift
should be consistent with that of the other neutral ions observed,
namely O~I and N~I.  We find that a single component fit allows for an
excellent description of the observed absorption, and we now consider the
individual parameters.

\subsubsection{Redshift}
The redshift of the hydrogen was determined by
simultaneously fitting the Lyman series absorption at \lya, \lyb,
\lyg, Ly-5, Ly-6, Ly-7, Ly-9, Ly-10, Ly-13, Ly-14, and Ly-15.

\subsubsection{Velocity Width}
The high column density of the absorber does
not give us information on the velocity width at \lya, so we must
turn to the higher order Lyman series lines.
To determine $b$, we fit the absorption in the Lyman series in
transitions which appear to be least contaminated by interloping \lya\
forest, specifically 
Ly-9, Ly-10, Ly-14, and Ly-15, simultaneously.  The inclusion of all
observed Lyman series transitions in the fitting procedure does not
changed the measured $b$ value.

\subsubsection{Column Density}
\begin{figure}
\centerline{\psfig{file=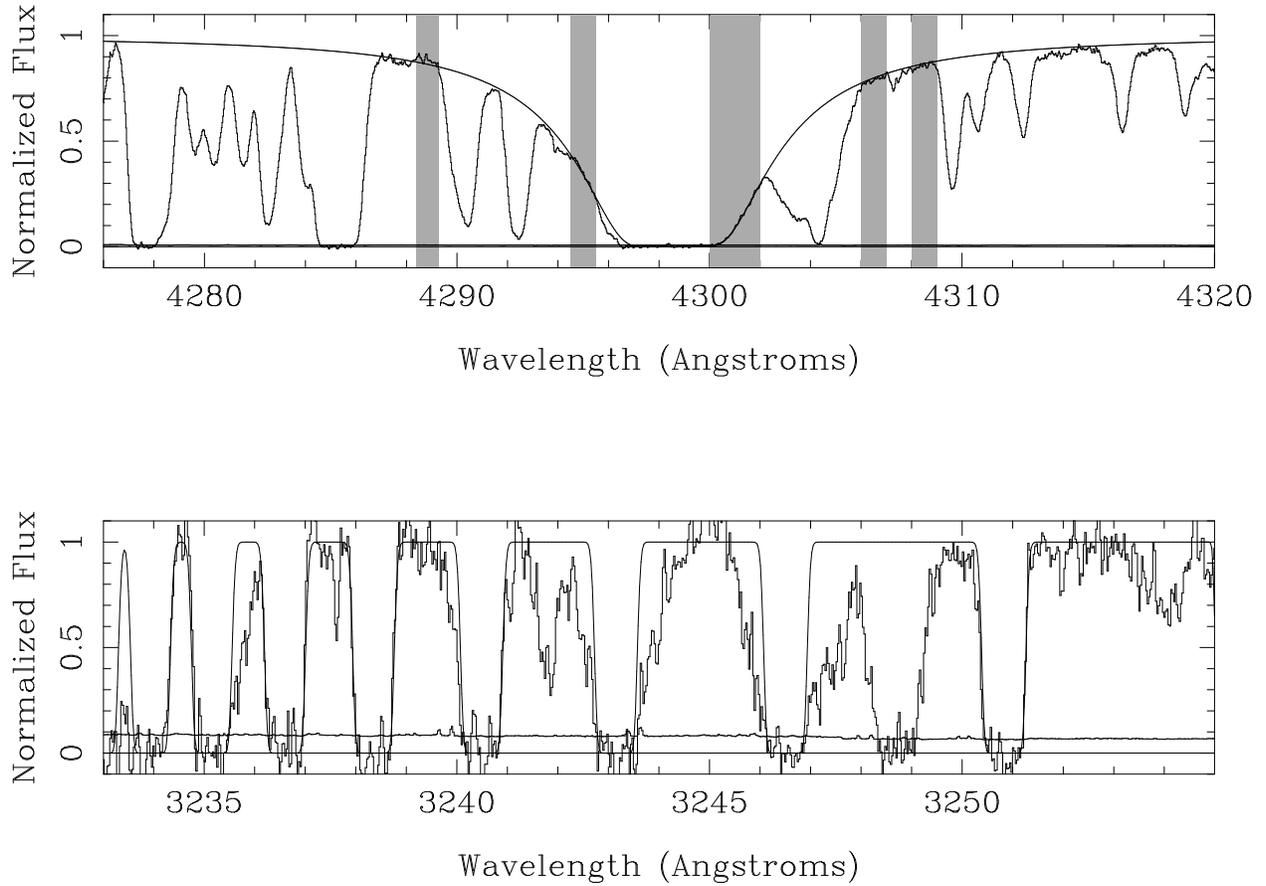, height=6.5in, angle=270}}
\caption{Spectral regions used to measure the H I column density.  The
upper panel shows the \lya\ line, with the core and damping wing
regions used in the fit shaded grey.
The lower panel shows the absorption near the
 Lyman limit. Over-layed is the single component fit to the hydrogen
with a column density of \lnhi $=19.422$ \cmm\ from Table
\ref{dhlinetab} .}
\label{alpha_limit}
\end{figure}

Three regions of the spectrum allow us to
determine the column density:  the Lyman limit, the core  of the
\lya, and the damping wings of the \lya.  These regions are shown in
Figure \ref{alpha_limit} along with the single component best fit to the H~I.    

The Lyman limit allows for a lower limit to the column density
of the hydrogen of \lnhi $=17.8$ \cmm\ since all observed Lyman series
lines observed are saturated.

\begin{figure}
\centerline{\psfig{file=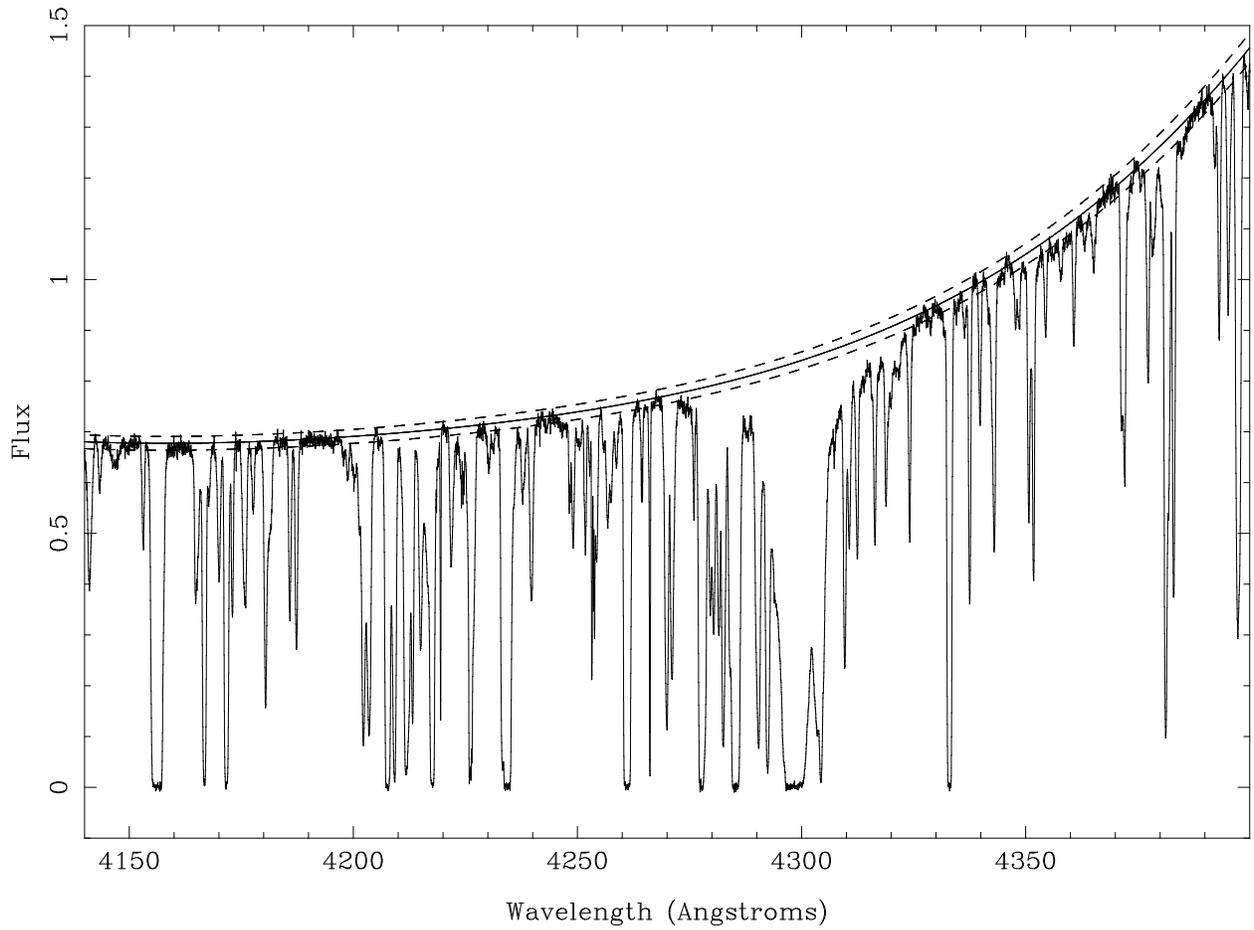, height=6.5in, angle=270}}
\caption{\lya\ region of the $z \simeq 2.536$ Lyman limit system.
Over-layed is the continuum fit (solid line) and the approximate $1
\sigma$ error to the continuum fit (dashed line).}
\label{contplot}
\end{figure}

The \lya\ line gives the most information regarding the hydrogen
column density, 
because it is insensitive to the $b$ value, even though it 
is sensitive to the continuum level.  In Figure
\ref{contplot}, we show the continuum used to fit the 250 \AA\ region
encompassing \lya\ and the corresponding approximate $1 \sigma$ levels, which
amount to a $\pm 2 \%$ continuum level change.  To determine the
effect of the continuum placement on the measured value on the column
density, for all regions of the \lya\ line, we obtain a fit using the
best continuum estimate, and then move the continuum to the $\pm 1
\sigma$ levels and re-fit.  In all fits to the \lya , we choose
segments of the spectra which
appear to be least contaminated by other absorption.

\begin{figure}
\centerline{\psfig{file=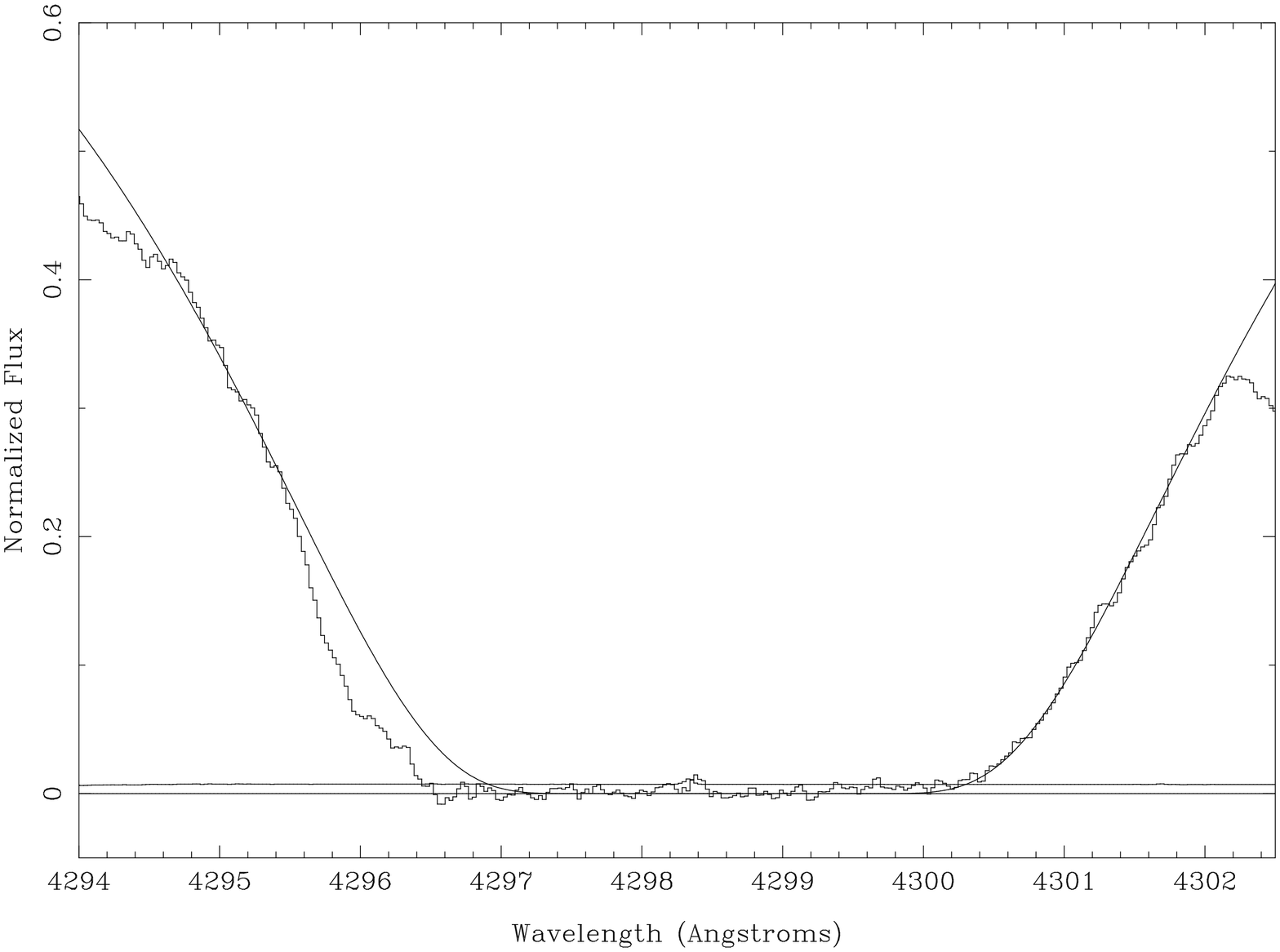, height=6.5in, angle=270}}
\caption{Fit to the core region of \lya\ with \lnhi $=19.419$ \cmm .}
\label{core_fit}
\end{figure}

In Figure \ref{core_fit}, we show the fit to the core of 
\lnhi $=19.419 \pm 0.009$ \cmm , where the error is the quadratic sum 
of the error from the continuum 
(0.007) and the error from the fit (0.006).  This fit was made
for two regions on either side of the line center:
4294.5--4295.5 and 4300--4302 \AA .
There is additional absorption in the core between 4295.5 and
4297 \AA\ which is readily fit by two H~I \lya\ lines.  Their
absorption is seen and fit in \lyb , but they contribute no
significant optical depth to the 
lower wavelength (4294.5--4295.5 \AA\ ) region which gives
the column density of the H~I which shows deuterium.

We note that
contamination of the core region is possible, and would lower the
measured column density.  However, we consider 
such contamination unlikely, since it
would require at least two lines appearing in the right places, with
a very restricted set of parameters, to 
produce enough absorption to fit both sides of the core region.

The damping wings of the \lya\ absorber also give
the column density, but the exact continuum placement is now the
dominant source of error because the continuum uncertainty represents
a larger fraction of the total absorption in these regions.  
Two damping wing regions were fit, one on
either side of the line center.  On the blue (lower wavelength) side,
we fit the wavelengths 4288.4--4289.3 \AA , which gave  \lnhi\ 
$= 19.406 \pm 0.060$ \cmm , where the errors are the range in
column density allowed by the $\pm 2\%$ range in the continuum
placement.
For the red (higher wavelength) side regions  
4306--4307 and 4308--4309 \AA\ were fit giving  \lnhi\  $= 19.4752
\pm 0.040$ \cmm .

For our best value for the H I column density, we take the weighted
mean of the three measurements
from the core and wings: \lnhi $=19.422
\pm 0.009$ \cmm .  All three estimates of the column
density are consistent with this mean. 

\subsection{The Deuterium}
The determination of the parameters describing the deuterium
absorption is relatively straightforward, given the presence of 5
un-blended transitions, most of which are unsaturated.  We begin by
fitting the D~I transitions, and then assess possible sources of
error.

\begin{figure}
\centerline{\psfig{file=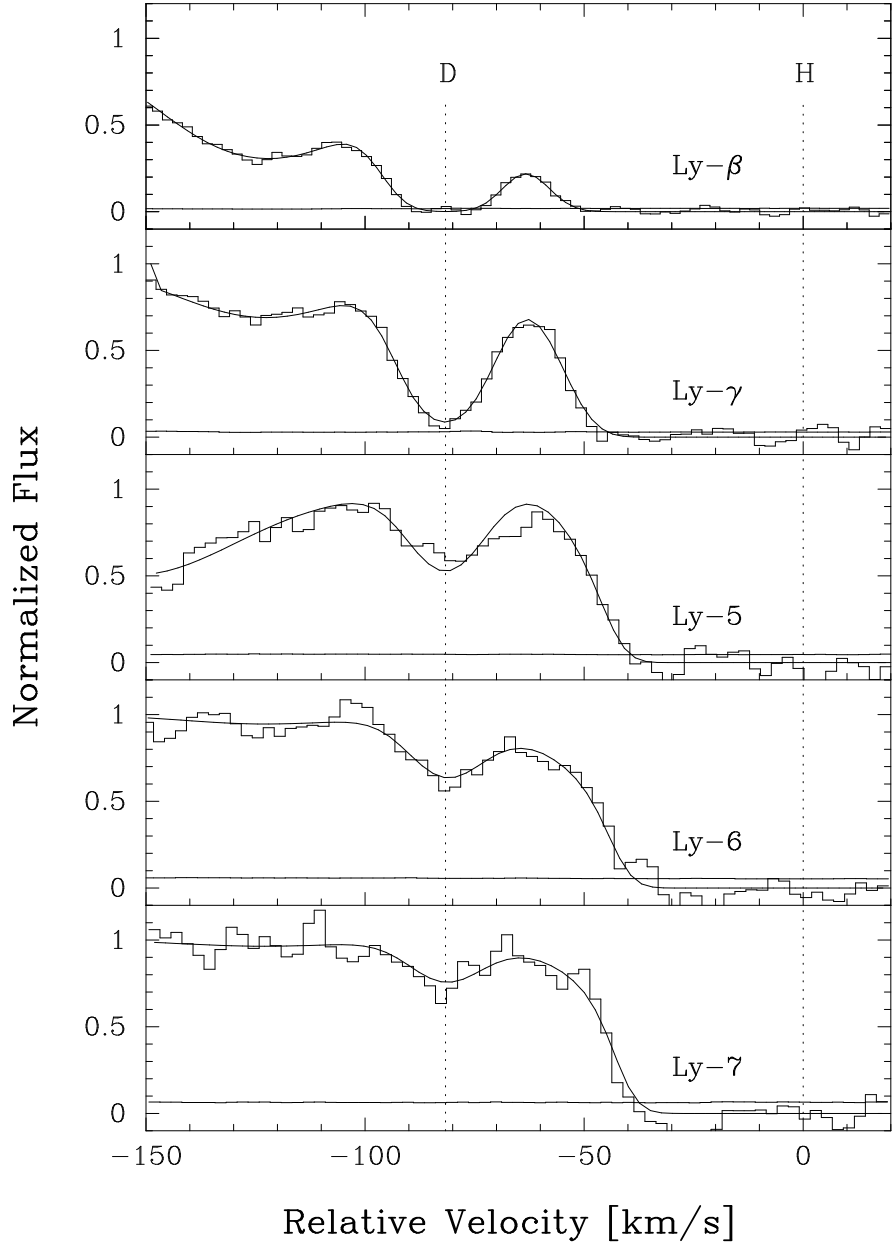, height=6.5in}}
\caption{Simultaneous fit to the deuterium at $-82$ \kms\ in 5 Lyman series
transitions with \lndi $=14.81$ \cmm\ and $b=9.93$ \kms . Also included in the
fit is the H at 0 \kms\ and additional \lya\ forest absorption.}
\label{dlinefit}
\end{figure}

The D~I transitions are well fit by an
absorber with \lndi$=14.810 \pm 0.029$ \cmm\ and a velocity width of $b =
9.93 \pm 0.29$ \kms.  This fit is shown in Figure \ref{dlinefit}.
These values were determined by fitting the regions
surrounding the deuterium in all 5 deuterium absorption regions simultaneously,
with the redshift of the deuterium tied to the hydrogen, 
and the parameters for the hydrogen absorption fixed at \lnhi$=19.42$ \cmm\ 
and $b=13.99$ \kms.  
In all fits, the regions used to fit deuterium were
constrained to be those within approximately
$-150$ and +20 \kms\ of the H~I line center, and are listed in Table
\ref{dregions}.
Where needed, additional \lya\
forest absorption was fit to model all absorption.
The parameters for the additional lines are given in table \ref{dlinestab}.

\begin{figure}
\centerline{\psfig{file=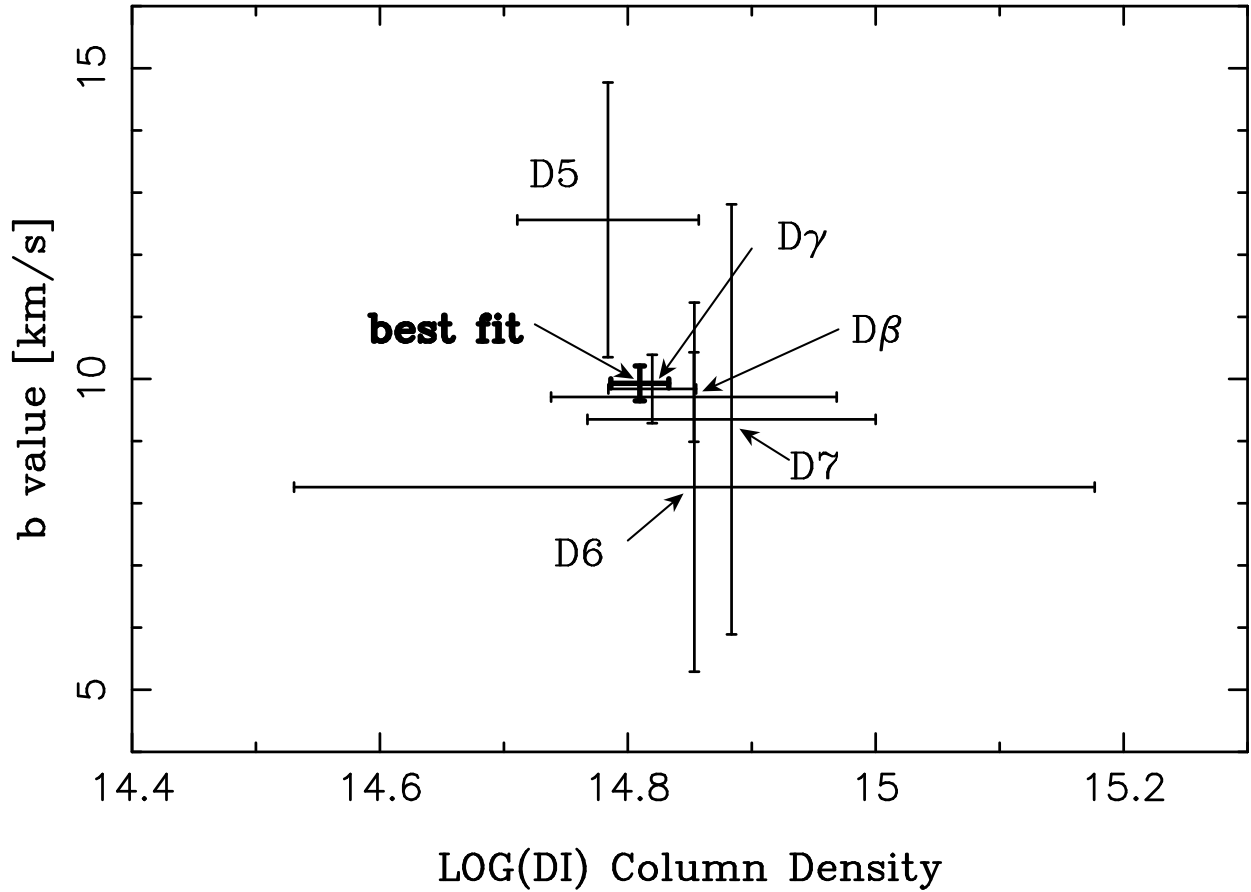, height=6.5in, angle=270}}
\caption{Values of the deuterium column density, \lndi\ (\cmm ), and the
absorption width parameter, $b$, for each of the 5 deuterium lines fit
separately.  The bold cross represents the values when all lines
are fit simultaneously using the best estimate for the continuum level.}
\label{dfits}
\end{figure}

The uncertainty in the continuum level increases the error on the D~I
column density by a third.  Independent estimates of the continuum
levels at the D~I transitions had a $1\sigma$ dispersion of 
approximately 10\% , much larger than the 2\% error near \lya , where
the data has been flux calibrated and has significantly higher SNR.
To gauge the effect on the error in the D~I column density,
 we determined the parameters of the deuterium
absorption independently for each line, both for the best estimate of
the continuum and for a 10\% higher continuum level.  The difference
between these two is the contribution to the error from the choice of
continuum level.  
For each each D~I transition, this error was added in quadrature to
the error obtained with the best continuum. The weighted mean for the
five D~I transitions gives 
\lndi $=14.826 \pm 0.039$ \cmm\ and $b = 9.85 \pm 0.42$\kms .  
The results of the independent fits to the different
deuterium transitions with the best estimate of the continuum level
can be seen in Figure \ref{dfits}.

The error on \lndi\ is insensitive to a change in the redshift of 1 \kms , 
as found by both letting the deuterium fit freely to its own redshift, 
and by tying the redshift to the H at 1 \kms\ away from
the best fit value.  Both results gave a \lndi\ which deviated 
by amounts significantly lower than the best fit $1\sigma$ errors.

The effect of varying the main hydrogen by the $1 \sigma$ errors in
either column density or velocity width produced no effective change
on the \lndi\ since the main hydrogen component is well separated from
the deuterium because the $b$ of the hydrogen is very small.

\subsection{The Metals}
\begin{figure}
\centerline{\psfig{file=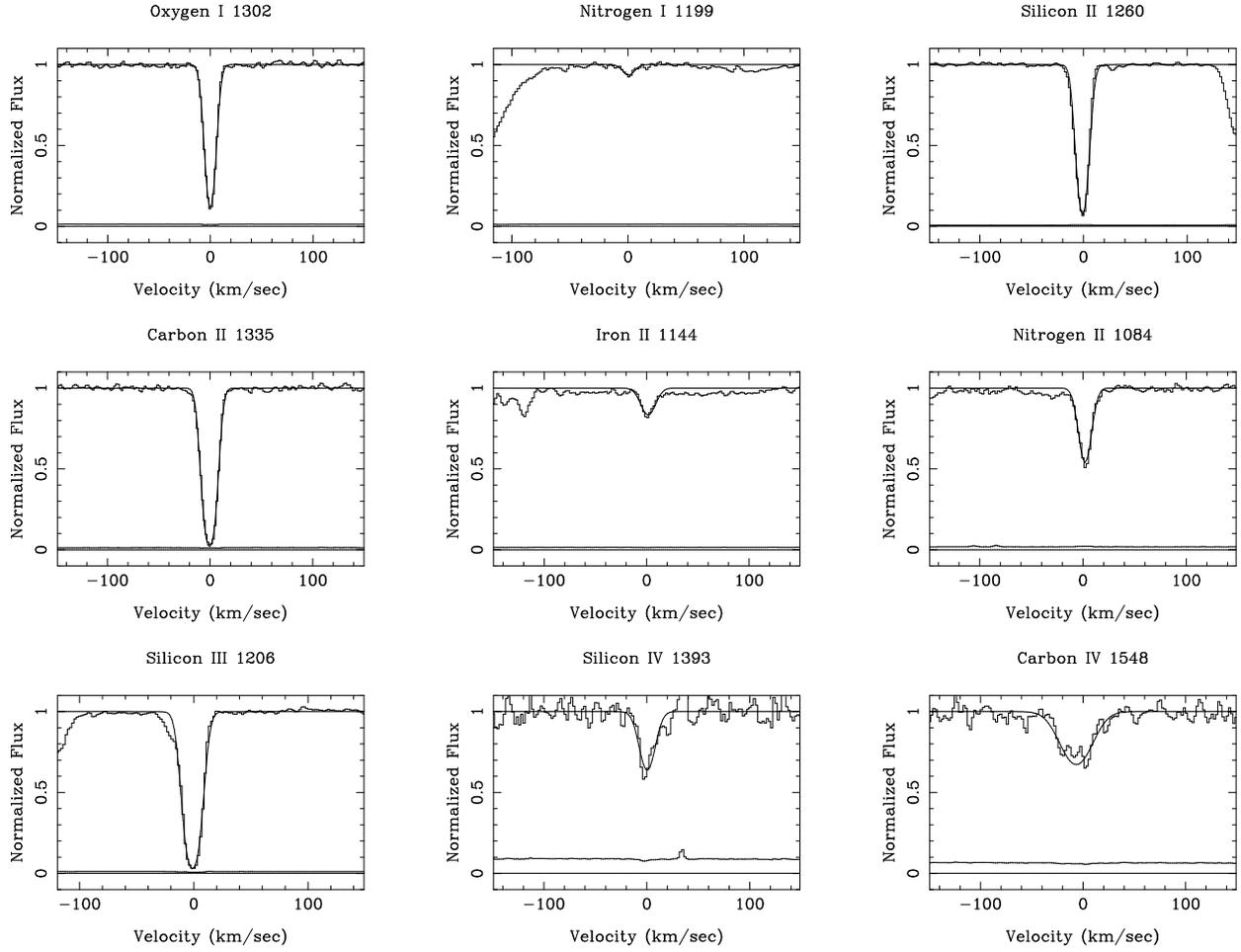, height=6.5in, angle=270}}
\caption{Best fit to metal lines associated with the $z \simeq 2.536$ LLS
towards \object.  The velocities shown are relative to $z = 2.535998$.}
\label{metalfit}
\end{figure}

The results of fits to the many metal lines are given in Table \ref{dhlinetab},
and a subset of the fits is shown in Figure \ref{metalfit}.
In all cases, the ions were best fit by a single component whose
redshift, velocity width, and column density were all allowed to vary.
Many of these ions show multiple transitions, and for any single ion,
all transitions present in the spectra were fit simultaneously.

The metals were all found to lie within 8 \kms\ of the redshift of the
hydrogen which shows deuterium.  
More importantly, since the column density of the
hydrogen is so high, we expect the redshifts of the low and neutral
 ions to agree with the hydrogen since we expect the gas to be
predominantly neutral.  The neutral and singly ionized ions
agree with the hydrogen redshift to within 2 \kms , while the neutral ions 
alone, in the top section of Table \ref{dhlinetab}, agree 
to within approximately 1 \kms .
The larger dispersion seen for the singly ionized ions, 
in the second section of the table, indicates that some of the gas
making these lines is distinct from the neutral gas, but this dispersion
could be insignificant, because the internal wavelength errors could be
1 --2 \kms .

The detection of O I is of particular importance in this system for two
reasons.  First, the ionization potential of O I is nearly identical
to that of hydrogen, so it is ionized by the same photons that ionize
the H I.  Second, O I participates in electron transfer with H I, such
that in cases where the gas is not highly ionized, O I/O is nearly
identical to H I/H, and the distribution of O I should match that of
the H~I and the D~I.  
The O I absorption gives information about the temperature, bulk motion,
ionization and abundances in the neutral gas which shows the D~I absorption.
Since the O I is accurately modeled by a single, narrow
component in four separate transitions, we gain confidence that a
single component fit to the deuterium and to the hydrogen is sufficient.
The implications of the measurement of O I on the
metallicity and ionization are discussed in a later section. 

Since O I and H I should arise in the same gas, we use these lines to 
obtain the temperature of the gas which shows D~I and 
its turbulent velocity.   We model the observed $b$ as
$b^{2} = b^{2}_{inst} + b^{2}_{int}$.  The instrument line
broadening, $b^{2}_{inst}$ was measured from arc lamp
calibration spectra to be $b_{inst}=4.81 \pm 0.14$~\kms. We model the
intrinsic velocity width as $b_{int}^2 = b_{temp}^2 + b_{turb}^2$; 
a combination of thermal broadening and bulk motion.
The thermal broadening,
$b_{temp}^2 = 2kT/m= 166.41(T/10^4 \rm K)/mass(amu)$, depends on 
the ion mass in atomic units, $m$, but the $b_{turb}$ is the same for all ions.
All of the $b$ values quoted in this
paper and listed in Table \ref{dhlinetab} refer to intrinsic line
widths, but the listed errors do not include the error in $b_{inst}$,
because we do not know whether this error is correlated at different
wavelengths.

Fitting O~I, N~I and H~I alone gives $T = 1.15 \pm 0.02 \times 10^4$ K, and
\bturb $= 2.56 \pm 0.12$~\kms , which we show by the straight line in
Figure~\ref{bplot}. The errors quoted here are very much minimum 
values, because they do not include the error in the $b_{inst}$ or the 
appropriateness of the model.

Ions C~II, Si~II and Fe~II, are all wider than predicted by this fit,
presumably because a part of each line arises in gas with different
velocity structure.  The lines from the higher ionization ions C~III,
C~IV, N~II, Si~III and Si~IV have velocities which differ by 0 to
--7~\kms\ from the H~I and low ionization ions. They are not relevant
to the D/H because their ionization and velocities show that they
arise in different gas, a common finding for absorption systems with
high \nhi\ (Wolfe \& Prochaska 2000a), but unlike absorption 
systems with much lower \nhi , including \qfirst\ and \qsecond .

\section{Is the Observed Absorption Deuterium?}
Now that we have determined the parameters of the absorption at the
position of deuterium, we turn to the issue of confirming that the
absorption is indeed deuterium, and not inter-loping hydrogen or metal line
contamination.

The primary concern with the absorption seen at the position of deuterium would
be that it is caused not by deuterium, but instead by inter-loping
hydrogen.  Here we argue that this scenario is unlikely for the
following reasons. 

\begin{figure}
\centerline{\psfig{file=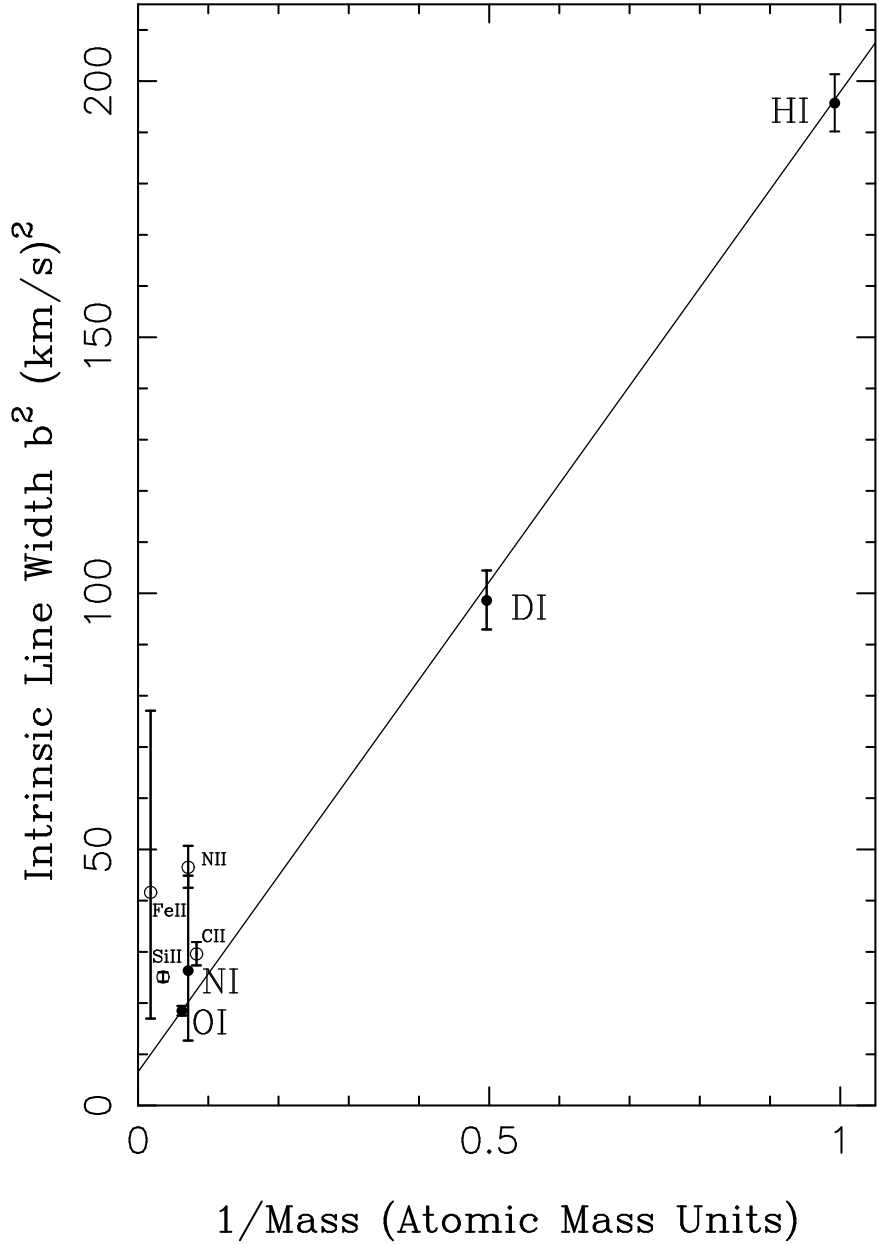, height=6.5in}}
\caption{Intrinsic velocity widths measured for the neutral and low
ionization ions.  The straight line is the best fit to H~I, N~I, and
O~I alone.  The slope of the line gives the temperature of the gas,
$T=1.15 \times 10^4$ K, and the intercept gives the turbulent
velocity, $b_{turb}=2.56$ \kms . The D~I absorption comes from the
same gas as the H~I, N~I and O~I.}
\label{bplot}
\end{figure}

First, hydrogen lines in the \lya\ forest with the appropriate column
density, $\log N_{HI} \simeq 14.8$ \cmm\ have $b > 20$ \kms , and not
$b \simeq 10$ \kms\ (Kim et al. 1997; Kirkman \& Tytler, 1997).  
However, such low values of $b$ might be found
in components of the LLS, which are the most likely contaminants.
Second, we are able to predict the width of the deuterium using the
measured widths of the other neutral ions observed which are present
in the same gas.
Figure \ref{bplot} illustrates the
concept.  Since we have observed three other neutral ions (H~I, O~I,
N~I), we can use their widths to predict the width of deuterium.  As Figure
\ref{bplot} shows, the measured value of $b(D)$ is consistent with its
predicted value.  
Third, hydrogen lines with a \lnhi $\simeq 14.8$ \cmm\ often show
associated metal line absorption, but as is seen in Figure
\ref{metalfit}, there is no such absorption at $-82$ \kms .
Finally, for the absorption to be hydrogen and not deuterium,
 its position would have
to agree with that of deuterium to within 1 \kms.  Taken together, these
arguments indicate that the observed absorption is deuterium, and not hydrogen, 
but
we can not quantify this because we do not know the properties of components
of Lyman limit systems.

The scenario in which the absorption is metal line contamination is
even less likely for a number of reasons.  Any metal lines
with a column density of $\log N_{metal} \simeq 14.8$ \cmm\ would show
absorption in not only that ion, but many others along with
strong associated hydrogen absorption, some of which would be easily
observed in our spectrum, but were not.  Also, for the observed
absorption to be entirely derived from metal line contamination, such
metal lines would have to appear in 5 different regions of the
spectrum, all at positions within 
approximately 1 \kms\ of the predicted positions of
deuterium, and with line strength scaling as the oscillator strengths
expected for the deuterium Lyman series.  A similar argument can be used to
exclude the case whereby the absorption at the position of deuterium was
hydrogen, but not in the corresponding Lyman series transition (e.g.,
the absorption at Ly-6 is an unrelated \lya\ line).

\section{Elemental Abundances and Ionization State of the $z \simeq 2.536$ Lyman
Limit System}

Here we discuss the abundances
of the elements observed in the $z \simeq 2.536$ Lyman limit system
which shows D/H.

The general procedure for determining the elemental abundances is to
use an ionization model to convert from the observed column densities
of selected ions into elemental abundances.  On the whole, the level
of ionization in QSO absorption line systems tends to decrease as the
column density of the gas increases, since the gas can shield itself
more from ionizing Lyman continuum radiation.  Typical Lyman limit
systems with \lnhi\ $\simeq 17.5$ \cmm\ are highly ionized, whereas
damped \lya\ systems, whose column densities are greater than
\lnhi\ $= 20$ \cmm\ are typically neutral.  In the case of
\object, we expect the gas to be predominantly neutral (H~I/H$ \ge 0.5$), 
since the column density is approaching that of a damped \lya\ system.

\subsection{Ionization \& Metallicity}

\begin{figure}
\centerline{\psfig{file=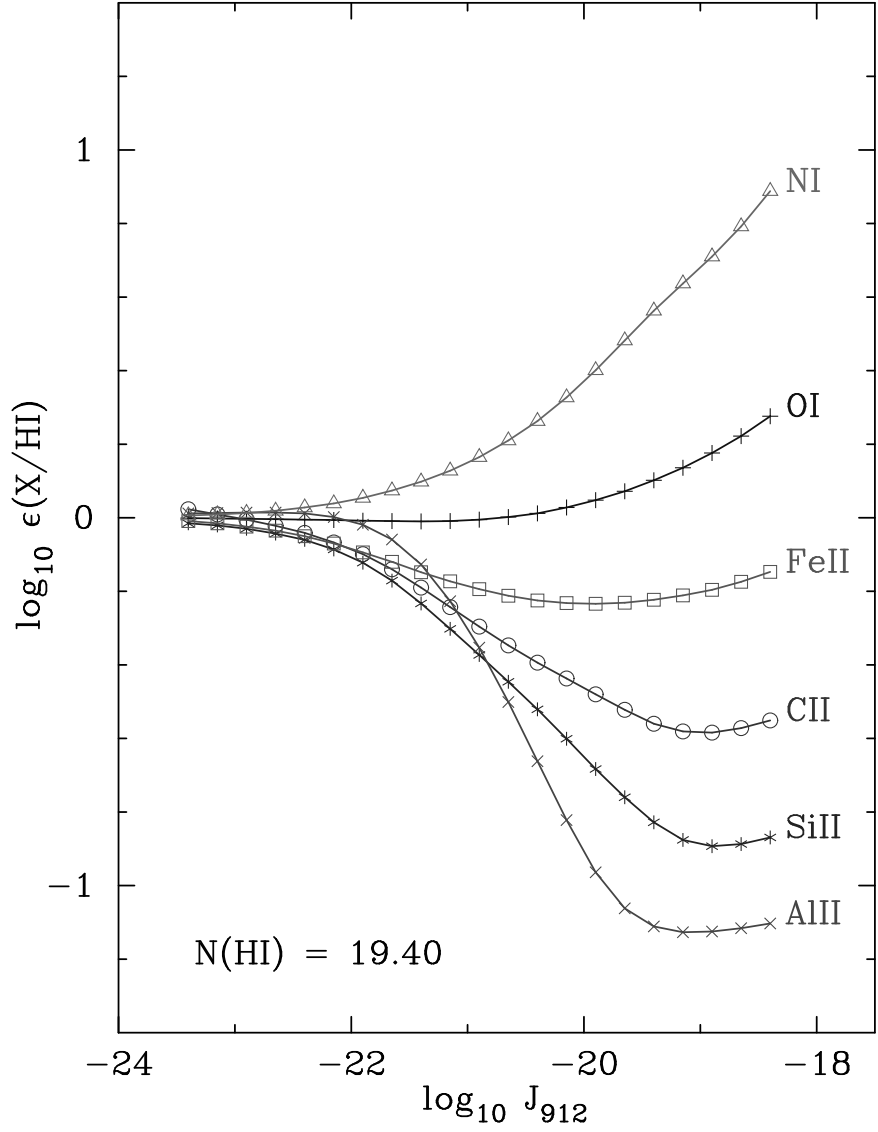, height=6.5in}}
\caption{
The predicted ionization correction $\log \epsilon ({\rm X/H~I})$ for
ion X as a function of the intensity of the photoionizing radiation,
$\log J_{912}$, in units of ergs cm$^{-2}$ s$^{-1}$ Hz$^{-1}$ sr$^{-1}$.
We conclude that the ionization is low, as on the left of the plot.
}
\label{figcldy}
\end{figure}

We modeled the ionization with the CLOUDY v90.04
package developed by G. Ferland (Ferland, 1991).  As usual, we assumed
a plane-parallel geometry with constant density $n_H = 0.01$ \cmmm,
and we approximated an isotropic background ionizing radiation
by placing a point source
at a very large distance.  We used the Haardt-Madau ionizing spectrum
(Haardt \& Madau, 1996).
In Figure~\ref{figcldy} we show the correction which would be added
onto an observed ionic abundance to
give an elemental abundance: 
[X/H] = $\log (N_{X}/N_{H~I}) + \log \epsilon ({\rm X/H~I}) -\log
(X/H)_{\odot}$, 
where $N_{X}$ is the column density of the ion used to infer the
abundance of element $X$.
Since we hold the \nhi , the gas density and the spectrum constant,
these corrections are specified in terms of the intensity of the
ionizing flux at 1~Ryd, $\log J_{912}$.
Because the corrections are nearly homologous with the ratio 
$n_{H}/J_{912}$, we can account for different $n_{H}$ values by
simply shifting $J_{912}$.

For the D/H absorber, we can obtain the [O/H] without
ionization
correction, because we have four un-blended O~I transitions which give
the most accurate $N_{OI}$ in any absorption system to date.
The ionization correction for O~I is negligible,
$\log \epsilon ({\rm O~I/H~I}) < 0.2$~dex, even at extreme values of
$J_{912}$, and O~I/H~I is equal to O/H, which gives [O/H] $= -2.0$.

This abundance is supported by four other elements.
Figure~\ref{figcldy} shows that the
ionization corrections for Si~II, C~II, Al~II, and Fe~II are all
negative.
Therefore, we can place upper limits on the C, Si, Fe, and Al
abundances using the observed
C~II/H~I, Si~II/H~I, Fe~II/H~I and Al~II/H~I ratios without
corrections.
We find [C/H] $\leq -1.9$, [Al/H] $\leq -2.1$, [Fe/H]$\leq -1.9$, and
[Si/H]$\leq -1.85$, all within 0.15~dex of [O/H].
Note that these elements might have shown different abundances,
 because they are seen in lines which have different velocities
and
velocity widths from the H~I, D~I and O~I.  From N~I, we find [N/H]
$\simeq -3.1$ without ionization correction, as suggested by the other
elements.  This implies that N is under-abundant by approximately 1.1
dex, which is not unusual for damped \lya\ systems with similarly low
metal abundances (Centurion et al. 1998; Lu, Sargent \& Barlow 1998),
but is much more N than usual in Galactic and extra-galactic H~II regions
(Henry, Edmunds \& K{$\rm\ddot o$}ppen 2000).

The similarity of these abundance values shows that each element is
primarily in the listed ion, O~I, Si~II, Fe~II, Al~II and C~II, and
that H~I/H $> 0.8$. This low ionization requires
$(J_{912}/10^{-21}$ erg \cmm\ s$^{-1}$ Hz$^{-1}$ sr$^{-1}$)
$(10^{-2}$\cmmm $/n_{H}) < 1$, and hence a weak radiation intensity and
high gas density. When we divide the \nhi\ by the gas density, we find
that size of the cloud along the line of sight would be
$l < 1$~kpc, assuming a homogeneous medium. If the absorbing region
were spherical, its gas mass would be $< 1.2 \times 10^5 M_{\odot} $.

The observed Si~III, C~III, C~IV or Si~IV absorption must come from
separate, more highly ionized gas.

\section{Best Fit Values for \object}

For the gas giving D/H in \object , the best values for various
parameters,
and the $1\sigma $ errors, are as follows:
\begin{itemize}
\item \lnhi $= 19.422 \pm 0.009$ \cmm\ (2\% error)
\item \lndi $=14.826 \pm 0.039$ \cmm\ (9\% error)
\item log D/H $= -4.596 \pm 0.040$ (10\% error)
\item D/H $=2.54 \pm 0.23 \times 10^{-5} $
\item temperature: $T = 1.15 \pm 0.02 \times 10^4$ K
\item Gaussian turbulent velocity width: \bturb $= 2.56 \pm 0.12$~\kms
\item log oxygen abundance, on the solar scale: [O/H] $= -2.0$
\item{other limits: [C/H] $\leq -1.9$, [Al/H] $\leq -2.1$, [Fe/H]
$\leq -1.9$, [Si/H] $\leq -1.85$, [N/H]$\simeq -3.1$}
\item neutral fraction: H~I/H$ > 0.8$
\item gas density: $n> 0.01$~\cmmm\
 (for $J_{912} \geq 10^{-21}$ erg \cmm\ s$^{-1}$ Hz$^{-1}$ sr$^{-1}$)
\item extent of absorbing region along the line of sight: $<1$ kpc
\item mass of gas: $<1.2 \times 10^5 M_{\odot}$. 
\end{itemize}
For D/H, we added the errors on \lnhi\ and \lndi\ in quadrature,
because we expect little correlation.

\section{Best values for D/H from all QSOs}

The weighted mean D/H from three QSOs,
\qfirst , \qsecond , and \object\ (the other QSO, \qthird , gives a
consistent upper limit) is 
$\log D/H = -4.523 \pm 0.026$ (6\% error).
The individual parameters from our measurements of four QSOs are
summarized in Table \ref{fourqsos}.
The dispersion in the three measurements of D/H is larger than
expected.  We obtain a $\chi^{2}_{2} = 7.1$, which would be exceeded in
3\% of random samples with Gaussian errors.  There are two interpretations.
Either we have under-estimated the errors, or there is real dispersion
in D/H.  We favor the former interpretation, since there are a
number of systematic errors which affect each measurement, and are
difficult to assess.  For example, we may have under-estimated the
effect of contamination in \qfirst\ or \qsecond.  Hydrogen
absorption near the D~I lines would artificially increase the D/H
estimated for these objects.  We now discuss another possible source of 
scatter in D/H, astration, the destruction of deuterium in stars.

\subsection{Primordial Nature of the Gas \& Astration}
At a metallicity of 0.01 solar, all chemical evolution models
which are consistent with Galactic data 
predict that the destruction of deuterium is about 1\%, ten times smaller than our
measurement error (Tosi et al. 1998; Jedamzik \& Fuller 1997).

We can get an approximate estimate of the amount of deuterium which will have been
destroyed in stars by linearly scaling from the local interstellar medium 
(LISM).
Since 50\% of the primordial deuterium is lost from the LISM where the metal
abundance is approximately 0.7 solar, we expect $0.7\%$ to be lost at 0.01
solar.  This scaling is only approximate for many reasons:
the relation is not linear in the LISM today
because the destruction is saturating,
there has been infall of less processed gas,
and the high mass stars which we expect polluted the
absorption systems destroy less deuterium for a given amount of metal
production.

\begin{figure}
\centerline{\psfig{file=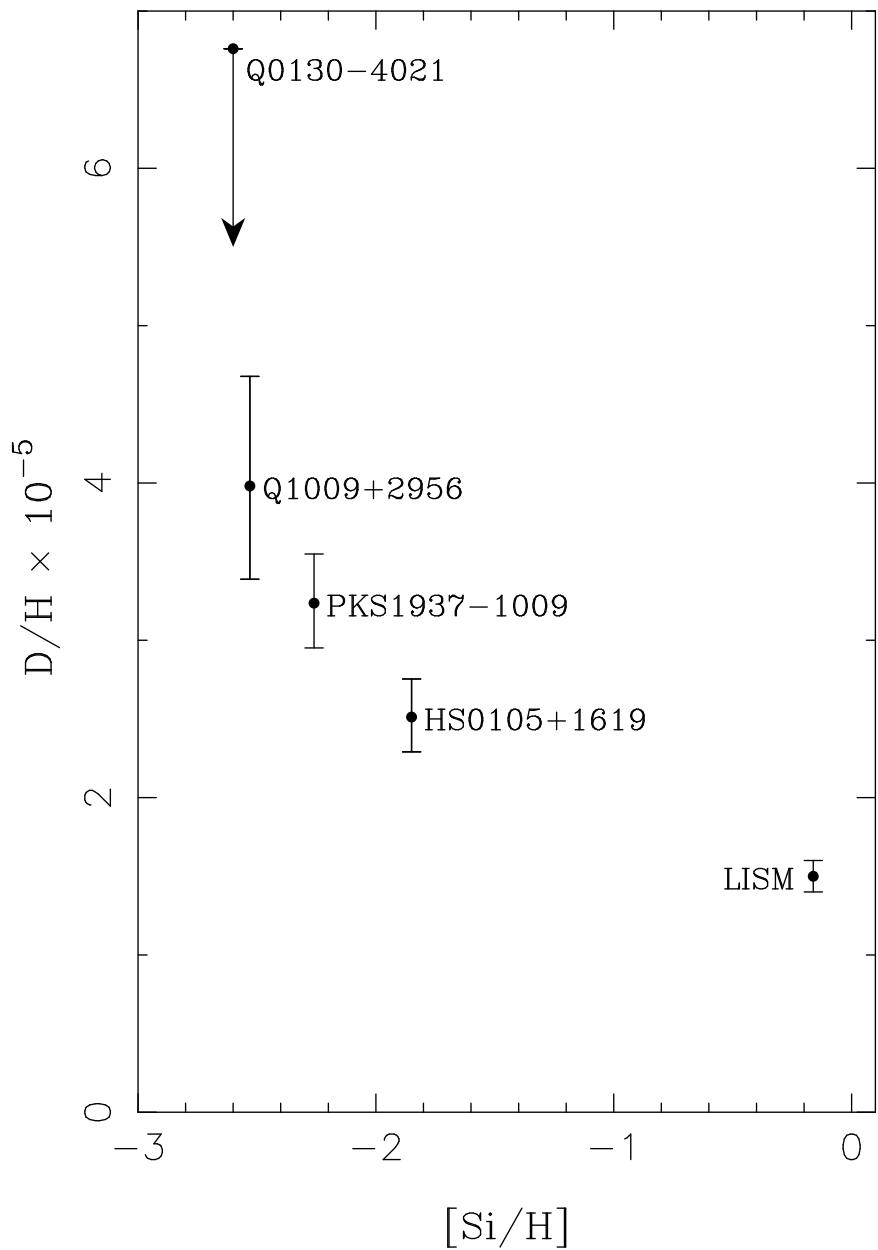, height=6.5in}}
\caption{
The relationship between D/H measurements and metal abundances, as gauged
by silicon. Standard chemical evolution models predict little change in
D/H until the [Si/H] $> -1$, at which time a significant fraction of
the gas has come from inside stars, where the D has been destroyed.  The
values of [Si/H] for \qfirst\ and \qsecond\ represent the mean value, 
weighted by the total H column density in each component:  $-2.26$ and $-2.53$
respectively.  The D/H value for the local interstellar medium (LISM) is
from Linsky (1998), and the LISM [Si/H] is from Savage \& Sembach
(1996).
}
\label{astration}
\end{figure}

We now discuss the relation between the D/H values and the metallicity
which we have measured towards QSOs.
The D/H values for the four QSOs listed in Table \ref{fourqsos} range from
$<6.7$ \mf\ to 2.54 \mf , while the [Si/H] ranges from $-2.6$ to $-1.85$.
Figure \ref{astration} shows that these values appear correlated, 
in the sense that the D/H is lower when the metallicity is higher.
This is the sense expected if the deuterium has been destroyed in those absorbers
with higher metallicity. 
The D/H for \qsecond , \qfirst\ and \object\ appears to decline linearly
with increasing [Si/H]. The D/H declines by a factor of 0.64 between
\qsecond\ and \object\ while the [Si/H] increases from approximately
$-2.53$ to $-1.85$.

There are two arguments against the reality of the correlation.
First, we do not expect such a large destruction of deuterium when the metal abundance 
remains this low. A decrease in the D/H by a factor of 0.64 implies that
0.64 of the atoms in the absorbing gas have not been in a star, and that
0.36 have been in one or more stars, which is at least an order
of magnitude more than expected.
Second, the indicated rate of decrease of the D/H is much faster than expected.
We expect D/H to decrease linearly with the linear, and not log, 
metal abundance.

Models of the evolution of the chemical abundances in our Galaxy
predict that D/H will remain near the primordial
value -- a horizontal straight line on Figure 12 -- for [Si/H] $< -1$
(Prantzos 1996, Fig. 4b;
Prantzos \& Silk 1998, Fig. 3;
Matteuci, Romano \& Molaro 1999, Fig. 7b).
The D/H declines at higher abundances, because a significant fraction of 
the atoms in the ISM have then come from within stars.
The abundance at which the D/H decline becomes significant depends on 
the details, including
the element used to gauge the metal abundance,
the amount of that element ejected from stars, 
the fraction of the H in the ISM which has not been inside a star, 
the infall of gas with abundances nearer to the primordial values,
and the amount of mixing.

The Galactic chemical evolution models apply on average to large volumes 
of gas which are well mixed. 
At the epoch of observation the gas may instead be isolated,
and the elemental abundances may have been determined by relatively few stars.
This will allow more stochastic variation in the D/H at a specific
metal abundance, but it will not change the general expectation that D/H
will remain near the primordial value as long as the metal abundances are low.

\section{New Values for the Relevant Cosmological Parameters}

We take the weighted mean of the D/H values from three QSOs,  
\qfirst, \qsecond\ and \object , as our best estimate for the primordial D/H 
ratio. We include all three QSOs because we believe that the dispersion in 
these D/H values is dominated by systematic errors. We weight by the quoted
errors, which we believe indicate the relative sizes of the random errors. 
For the error on the best value of D/H we take the dispersion in the 
three values, divided by $\sqrt{3}$. We do not use the error on the weighted 
mean, which is unrealistically small, 
because the dispersion on the values is larger than
expected. We then estimate:
\begin{itemize}
\item log D/H $= -4.52 \pm 0.06$ (15\% error), or
\item D/H  $= 3.0 \pm 0.4 $\mf .
\end{itemize}
This new D/H value,
together with over 50 years of theoretical work and laboratory measurements
of reaction rates, leads to the following values for cosmological
parameters (Esposito et al 2000a,b; Burles, Nollett \& Turner, 2000):
\begin{itemize}

\item $\eta = 5.6 \pm 0.5 \times 10^{-10} $ (from standard BBN and D/H)
\item \ob $= 0.0205 \pm 0.0018$ (baryon density, in units of
the critical density, 9\% error)
\item $Y_p = 0.2471 \pm 0.0009$ (predicted mass fraction 
of $^4$He, from SBBN and D/H)
\item $\rm ^3He/H = 1.09 \pm 0.06 \times 10^{-5}$ (predicted from SBBN
and D/H)
\item $\rm ^7Li/H = 3.8^{+1.0}_{-0.8}  \times 10^{-10}$ (predicted from SBBN
and D/H),
\end{itemize}
where the error on $Y_p$ includes the error from D/H (0.0008) and from
the calculation of $Y_p$ for a given $\eta$ (0.0004) (Lopez \& Turner
1999).

These values are generally consistent with most other measurements
(Tytler et al. 2000; Burles, Nollett \& Turner 2000; Olive, Steigman
\& Walker 2000).  For example, Pagel (2000) reviews \yp\ measurements and 
concludes that ``systematic errors up to about 0.005 are still not excluded",
and that ``\yp\ is very probably between 0.24 and 0.25''.
Of special note are $^7$Li and the cosmic microwave
background (CMB) estimate of \ob .

The predicted value for the $^7$Li/H abundance is a factor of
two to five times higher than measured in halo stars 
(Bonifacio \& Molaro 1997; Ryan, Norris \& Beers 1999). Models 
allow at most a factor of two destruction of the $^7$Li in these stars
(Ryan et al. 1999, 2000; Deliyannis \& Ryan 2000; Tytler et al. 2000).

Recent measurements of \ob\ from the CMB are in agreement with
our value from D/H and SBBN. Compared to our more accurate
\ob\ value, the value from
BOOMERANG (Lange et al. 2000: \ob $=0.031 \pm 0.004$) is approximately
$3\sigma$ higher,
that from 
MAXIMA-I (Balbi et al. 2000: \ob $= 0.025 \pm 0.010$) is $0.5\sigma$ higher
and that from the Cosmic Background Imager (Padin et al. 2001: \ob $= 0.009$)
is just over $1\sigma $ lower.
The agreement of these \ob\ values is a dramatic validation of the physics 
used in SBBN and CMB.

\section{Summary \& Discussion}
We have presented the fourth quasar, \object , which shows an
absorption system having a low deuterium to hydrogen abundance ratio:
D/H$ = 2.54 \pm 0.23 \times 10^{-5}$ 
in the $z \simeq 2.536$ Lyman limit absorption
system.  We first obtained low resolution spectra in our survey for
D/H QSOs with the Lick 3 meter Shane telescope, and here we presented
over 24 hours of spectra from the HIRES spectrograph on the Keck-I
telescope.

The absorber has a high neutral hydrogen column density, \lnhi $=19.422 \pm
0.009$ \cmm\ which is 36 times larger than the next highest case studied for
D/H, but 8--100 times less than standard damped Lyman alpha systems.
Very little is known about absorbers with \lnhi $\simeq 19$ \cmm , and this
absorber may not be representative of this class because it was selected
to have a very simple velocity structure and low $b$ values.

While the absorber towards object has by far the highest neutral H~I column
density of the absorbers which we have studied for D/H, it
also has the lowest total Hydrogen column density, 
by a factor of 4 -- 7, when we correct for the ionization.

We know little about the environment around the absorber.
It might be in the outer parts of galaxy, as are LLS at low redshift, or 
in a disk, as are damped Lyman alpha absorbers at high redshift.
Alternatively, it may be in a relatively isolated gas cloud, because
see just one component, with an exceptionally small spread of velocities.
In either case, numerical simulations of
the growth of structure suggest that the absorbing gas 
has been incorporated into a galaxy by today. 

For the first time, we detect deuterium absorption in 5 Lyman series
transitions, and determine \lndi $=14.826 \pm 0.039$ \cmm .  We have strong
arguments that the observed absorption is indeed deuterium, and not
inter-loping hydrogen or metal line absorption.

We observe a number of associated metal line absorbers, from which we
calculate that the gas is warm and neutral.
The metallicity of the system is $\simeq 0.01$ times
solar, indicating that the measured D/H is
representative of primordial D/H.

We argue that \object\ offers the most secure
detection of D/H to date, and that the D/H ratio determined from all
QSOs has been made more secure.

\subsection{\object\ gives the most secure measurement of primordial
D/H}

The measurement of primordial D/H towards \object\ is more secure than
our prior measurements towards \qfirst , \qsecond\ and \qthird\ for
several reasons. By secure we mean that we have the most information, and
hence there
is less chance of undetected errors which might greatly exceed those
quoted.
We are not explicitly referring to the size of the quoted errors, which
are similar for \qfirst\ and \object , and larger for \qsecond\ .

The absorption system in \object\ is simple.
Like \qthird , the second most secure result, the absorber in \object\
is modeled with a single component, which simplifies the measurement 
of the column densities.
\qfirst\ was modeled with two or three components, and \qsecond\ with
two to four.

\object\ is the only case to show more than one strong deuterium line, which
reduces the chance of contamination, and gives more reliable $b$
values, and hence \ndi .

For \object\ we listed above the many reasons why the absorption near
the deuterium line position is deuterium. The chance of serious
contamination from the \lya\ forest will decrease with rising \ndi .
Such contamination is least likely to be significant in \object ,
followed by \qfirst . There is some contamination in \qsecond\ while for
\qthird\
we see a lot of contamination, and obtain only an upper limit on \ndi .
We do not know whether the chance of contamination by components of
the Lyman limit system changes with \nhi .

We also expect that the chance of contamination decreases as the $b$
value of the deuterium decreases, because H lines often have larger $b$ values.
Hence \object\ is the most secure detection of deuterium.

For \object\ we have the most information on the velocity field and
$b$ values because we see several D~I, N~I and O~I lines.

For \object\ the metal abundance is obtained with additional
redundancy.
The gas is nearly neutral, and hence we get the abundances of
several elements: H, C, N, O, Al, Fe and Si.
For the \qfirst\ and \qsecond\ we used a standard photoionization
model to find the level of ionization which explained the relative abundances
of ions such as C~II, C~III, C~IV and Si~II, Si~III, Si~IV. We obtained a
solution for each element, and these agreed, which provided a check.
For \qthird\ the ionization and metal abundances are both less well
known.

\subsection{The primordial D/H becomes more secure}

The new measurement makes the primordial D/H much more secure because
in each case we are sampling gas with different physical conditions, and
some systematic errors, including those associated with the
measurement of column densities, may be different for each QSO.
First, the new measurement is the most secure.
Second, we have increased the number of QSOs in which we have measured
deuterium from two to three.
Third, we sample a new region of space.
Each absorber is a different direction in the universe, and each
samples a sight line of about 1 -- 10~kpc, which requires about
10$^5 - 10^6$ solar masses of gas.
Fourth, the absorbing gas covers a factor of 240 range in \nhi .
Fifth, the absorption systems have differing ionization, with a range
of 2000 in the H~I/H ratio.

There is less variation in other parameters. The metal abundances cover
a factor of ten, which is the typical for QSO absorption line systems,
while the redshifts cover most of the range observable from the
ground.

\section{Acknowledgments}

This work was funded in part by grant G-NASA/NAG5-3237 and by NSF
grants AST-9420443 and AST-9900842. The flux calibration was funded in
part by NAG5-9224.

The spectra were obtained from the Lick Observatory, and the
W.M. Keck observatory, which is managed by a partnership among
the University of California, Caltech and NASA.

D.T. thanks Elizabeth Flam for hosting a visit to the Institut de
Astrophysique in Paris.

We are grateful to Steve Vogt, the PI for the Keck
HIRES instrument which enabled our work on D/H, and to the W.M. Keck
Observatory staff; the observing assistants
Joel Aycock, Teresa Chelminiak, Gary Puniwai, Ron Quick, Barbara
Scheafer, Cynthia Wilburn,
Chuck Sorenson, Terry Stickel and Wayne Wack,
and the instrument specialists
Tom Bida, Randy Campbell, Bob Goodrich, David Sprayberry and Greg
Wirth.
We are especially grateful to H.J. Hagen, D. Engels and D. Reimers for
sending us information on this QSO prior to publication.
We also thank 
Scott Burles, 
Salvatore Esposito, 
Craig Hogan,
Sergei Levshakov,
Gianpiero Mangano,
Gennaro Miele, 
Ofelia Pisanti,
Tony Readhead,
Gary Steigman,
Mike Turner
and
Ned Wright 
for useful discussion and comments.


\begin{deluxetable}{llcccc}
\tablecaption{\label{obstab}OBSERVATIONS OF \object}

\tablehead{
\colhead{Instrument} &
\colhead{Date} &
\colhead{Integration Time} &
\colhead{Wavelengths covered}
\\
\colhead{} &  \colhead{} & \colhead{(seconds)} &
\colhead{(\AA)}
}

\startdata
Kast & August 16, 1998 & 2700 & 3200 -- 5100 \nl
HIRES & October 10, 1999 & 7200 & 3200 -- 4720 \nl
HIRES & October 11, 1999 & $2 \times 8000$ & 3200 -- 4720 \nl
HIRES & November 9, 1999 & 1800 & 4220 -- 6640 \tablenotemark{a} \nl
HIRES & September 19, 2000 & $2 \times 7200$ & 3200 -- 4720 \nl
HIRES & September 20, 2000 & 8600, 10800 & 3200 -- 4720 \nl
HIRES & September 20, 2000 & $4 \times 7000$ & 3200 -- 4720 \nl
Total (HIRES): & & 86,800 & \nl

\enddata
\tablenotetext{a}{\footnotesize{The wavelength coverage for this
observation was not continuous due to 1-10 \AA\ spectral order gaps.}}
\end{deluxetable}

\begin{deluxetable}{lcrcc}
\tablecaption{\label{dhlinetab} IONS OBSERVED IN
        THE $z \simeq 2.536$ LLS TOWARDS \object \tablenotemark{a}}
\tablewidth{34pc}

\tablehead{
\colhead{Ion} &
\colhead{log N} &
\colhead{$b$\tablenotemark{b}} &
\colhead{$z$} &
\colhead{$v$\tablenotemark{c}}
\\
\colhead{} & \colhead{(\cmm)} & \colhead{(\kms)} & \colhead{} & \colhead{(\kms )}
}


\startdata
H I 
& $19.422 \pm 0.009$\tablenotemark{d} 
& $13.99 \pm 0.20$ & $2.535998 \pm 0.000007$ 
& $~~0.0 \pm 0.6$\tablenotemark{e} \nl

D I
& $14.826 \pm 0.039$\tablenotemark{f} 
& $9.85 \pm 0.42$\tablenotemark{f} & $2.536002\pm 0.000008$ 
& $0.4 \pm 0.7$ \nl

N~I
& $12.306 \pm 0.060$ & $5.13 \pm 1.57$ & $2.535998 \pm 0.000009$
& $0.0 \pm 0.8$\nl

O I
& $14.378 \pm 0.024$ & $4.30 \pm 0.11$  & $2.535991 \pm 0.000001$ 
& $-0.6 \pm 0.1$ \nl

\hline

C II
& $14.349 \pm 0.069$ & $5.44 \pm 0.21$  & $2.535985 \pm 0.000001$ 
& $-1.1 \pm 0.1$ \nl

Si II
& $13.156 \pm 0.012$ & $5.01 \pm 0.09$  & $2.535980 \pm 0.000001$ 
& $-1.5 \pm 0.1$ \nl

Al~II
& $11.870 \pm 0.040$ & $9.69 \pm 0.99$ & $2.536019 \pm 0.000009 $
& $+1.8 \pm 0.8$ \nl

Fe II
& $12.913 \pm 0.087$ & $6.45 \pm 2.33$  & $2.535983 \pm 0.000014$
& $-1.3 \pm 1.2$ \nl

\hline

C III
& $13.716 \pm 0.032$ & $14.48 \pm 0.56$  & $2.535923 \pm 0.000004$ 
& $-6.4 \pm 0.4$ \nl

C IV\tablenotemark{g}
& $13.277 \pm 0.019$ & $20.40 \pm 1.07$  & $2.535913 \pm 0.000009$ 
& $-7.2 \pm 0.8$ \nl

N II
& $13.501 \pm 0.013$ & $6.82 \pm 0.30$  & $2.536015 \pm 0.000002$ 
& $+1.4 \pm 0.2$ \nl

Si III
& $13.097 \pm 0.037$ & $7.10 \pm 0.24$  & $2.535977 \pm 0.000001$ 
& $-1.8 \pm 0.1$ \nl

Si IV\tablenotemark{g}
& $12.647 \pm 0.053$ & $9.16 \pm 1.59$  & $2.535995 \pm 0.000014$
& $-0.3\pm 1.2$ \nl

\enddata
\tablenotetext{a}{\footnotesize{
The three sections group ions by increasing ionization.
For ions in the same gas, we expect the $b$ values to decrease 
with increasing mass.
Errors quoted in the table for the N and $z$ values are from VPFIT alone, 
except for H~I and D~I.
}
}
\tablenotetext{b}{\footnotesize{
The intrinsic $b$ value.
}
}
\tablenotetext{c}{\footnotesize{
Velocities are all relative to $z=2.535998$.
The errors in $v$ values come from the errors listed on the $z$ values: 
$\sigma(v)^2 = (c\sigma(z)/(1+z))^2 + 0.09^2$, 
where 0.09~\kms\ is the minimum internal uncertainty in the wavelength scale. 
The internal error may be 1 -- 2 \kms , while the
external error is approximately $\pm 10$ \kms .
}
}
\tablenotetext{d}{\footnotesize{
The value for \lnhi\ is the weighted mean from the damping wings and
core region of \lya, and the error includes the continuum level error.
}
}
\tablenotetext{e}{\footnotesize{
The H~I $z$ defines $v=0$. The 0.6 \kms\ error is the
uncertainty in the $v$ of the H lines in this frame.}
}
\tablenotetext{f}{\footnotesize{
The $N$ and $b$ for the D~I are the weighted means of the individual
fits to the five D transitions, and the error on the $N$ includes the
contribution from the continuum uncertainty.}
}

\tablenotetext{g}{\footnotesize{
The lines of this ion may be multiple component blends.}
}
\end{deluxetable}

\begin{deluxetable}{llcc}
\tablecaption{\label{dregions}SPECTRAL REGIONS USED TO MEASURE D~I}
\tablehead{
\colhead{Region} &
\colhead{$\lambda_{min}$ (\AA )} &
\colhead{$\lambda_{max}$ (\AA )} &
}
\startdata
\lyb & 3625.1 & 3627.4 \nl
\lyg & 3437.2 & 3439.2 \nl
Ly-5 & 3314.2 & 3316.4 \nl
Ly-6 & 3289.4 & 3291.4 \nl
Ly-7 & 3273.5 & 3275.2\nl

\enddata
\end{deluxetable}

\begin{deluxetable}{llccc}
\tablecaption{\label{dlinestab}ADDITIONAL LINES USED TO FIT D REGIONS}
\tablehead{
\colhead{$\log N$ (\cmm )} &
\colhead{$b$ (\kms )} &
\colhead{$z$} &
\colhead{Region\tablenotemark{a}} &
}
\startdata
14.324 & 25.2 & 2.53455 & all\tablenotemark{b} \nl
14.901 & 18.2 & 2.53554 & all\tablenotemark{b} \nl
13.365 & 26.2 & 1.72637 & Ly-5 \nl
12.944 & 29.7 & 1.70682 & Ly-6 \nl
12.770 & 39.7 & 1.69376 & Ly-7 \nl
\enddata
\tablenotetext{a}{\footnotesize{The regions are defined in Table
\ref{dregions}}}
\tablenotetext{b}{\footnotesize{All D regions used:
\lya , \lyb , \lyg, Ly-5, Ly-6, and Ly-7}}
\end{deluxetable}

\begin{deluxetable}{lcccccc}
\tablecaption{\label{fourqsos} PARAMETERS FOR THE D/H MEASUREMENTS}
\tablewidth{42pc}
\tablehead{
\colhead{QSO} &
\colhead{$\log $D/H} &
\colhead{$z_{dh}$} &
\colhead{\lnhi\ } &
\colhead{$\log n_{H~I}/n_H$} &
\colhead{$b$(D)} &
\colhead{[Si/H]}
\\
\colhead{} & \colhead{} & \colhead{} & \colhead{(\cmm )} & \colhead{} &
\colhead{(\kms )}
& \colhead{}
}
\startdata
\qfirst\ \tablenotemark{a}
& $-4.49 \pm 0.04$ & 3.572 & $17.86 \pm 0.02$ & $-2.35,-2.29$ & $14.0 \pm 1.0$ &
$-2.7,-1.9$\nl
\qsecond\ \tablenotemark{b}
& $-4.40 \pm 0.07$ & 2.504 & $17.39 \pm 0.06$ & $-2.97,-2.84$ & $15.7 \pm 2.1$ &
$-2.4,-2.7$\nl
\qthird\ \tablenotemark{c}
& $< -4.17$ & 2.799 & $16.66 \pm 0.02$ & $-3.4$ & $16 - 23$ & $-2.6$\nl
\object\ \tablenotemark{d}
& $-4.60 \pm 0.04$ & 2.536 & $19.422 \pm 0.009$ & $-0.1$ & $9.85 \pm 0.42$ & 
$-1.85$\nl
\enddata
\tablenotetext{a}{\footnotesize{
We list the parameters for each of the two components, where
available. Results
from Tytler, Fan \&\ Burles (1996); Burles \&\ Tytler (1998a).}
}
\tablenotetext{b}{\footnotesize{
We list the parameters for each of the two components, where
available. Results from Tytler \& Burles (1997) 
and Burles \&\ Tytler (1998b).} }
\tablenotetext{c}{\footnotesize{From Kirkman et al. (1999).} }
\tablenotetext{d}{\footnotesize{This paper.} }
\end{deluxetable}

\begin{thebibliography}{}

\bibitem{adams76}
Adams, T.F., Astron. Astrophys. {\bf50}, 461, (1976).
\bibitem{1937nhi}
Burles, S. \& Tytler, D., Astron. J. {\bf114}, 1330
astro-ph/9707176 (1997).

\bibitem{balbi_MAXIMA}
Balbi, A. et al. 2000, preprint, astro-ph/0005124

\bibitem{bonif_li}
Bonifacio, P. \& Molaro, P. 1997, MNRAS,{\bf285}, 847

\bibitem[Burles \& Tytler 1998a 1937]{bur98a}
Burles, S. \& Tytler, D. 1998a, ApJ, {\bf499}, 699

\bibitem[Burles \& Tytler 1998b 1009]{bur98b}
Burles, S. \& Tytler, D. 1998b, ApJ,{\bf 507}, 732

\bibitem{bur2000a}
Burles, S., Nollett, K. \& Turner, M. 2000, preprint astro-ph/0008495

\bibitem{bur2000b}
Burles, S., Nollett, K. \& Turner, M. 2000, preprint astro-ph/0010171

\bibitem{cent_nitrogen}
Centurion, M., Bonifacio, P., Molaro, P. \& Vladilo, G. 1998, ApJ
{\bf509}, 620

\bibitem{li6 measures}
Deliyannis, C.P. \& Ryan, S.G. 2000 Bull AAS 196.0602

\bibitem{esp2000a}
Esposito, S., Mangano, G., Miele, G. \& Pisanti, O. 2000a, Nucl. Phys.
B {\bf568}, 421

\bibitem{esp2000b}
Esposito, S., Mangano, G., Miele, G. \& Pisanti, O. 2000b, 
astro-ph/0005571

\bibitem[Ferland (1991)]{fer91}         
Ferland, G. J. 1991, Ohio State Internal Report 91-01

\bibitem[Haardt \& Madau (1996)]{haa96}
Haardt, F. \& Madau, P. 1996, \apj, {\bf 461}, 20

\bibitem[]{hagen99} Hagen, H.J., Engels, D., \& Reimers, D. 1999,
A\&AS, {\bf 134}, 483H

\bibitem{henry00} Henry, R.B.C., Edmunds, M.C., K{$\rm\ddot o$}ppen , J. 2000
ApJ {541} 660

\bibitem{jaffe_boom}
Jaffe, A.H. et al. 2000, preprint, astro-ph/0007333

\bibitem[]{is d primordial}
Jedamzik, K. \& Fuller, G.M. 1997, ApJ, {\bf483}, 560

\bibitem[]{kim97} Kim, T-S., Hu, E.M., Cowie, L.L. \& Songaila, A. AJ
{\bf 114}, 2 (1997)

\bibitem[]{kirkman97a} Kirkman, D., and Tytler, D. 1997, ApJ, {\bf
484}, 672

\bibitem[]{kirkman99} Kirkman, D., Tytler, D., Burles, S., Lubin, D.,
\& O'Meara, J.M. ApJ {\bf 529}, 655--660, (2000)
astro-ph/9907128

\bibitem{kt90}
Kolb, E.W., and Turner, M.S., ``The Early Universe'', (Addison Wesley
1990).

\bibitem{lange_ob}
Lange, A.E. et al. 2000, preprint, astro-ph/0005004

\bibitem{lev1718}
Levshakov, S.A., Kegel, W.H., \& Takahara, F., Astron. Astrophys.,
{\bf336}, L29 astro-ph/9801108 (1998).

\bibitem{levnewmethod}
Levshakov, S.A., Agafonova, I.I., Kegel, W.H. 2000, A\&A, {\bf 360}, 833-845
astro-ph/0003078

\bibitem[]{levsh-molaro_a} Levshakov, S.A., Agafonova, I.I., \& Kegel,
W.H. 2000, A\&A {\bf 355}, L1 astro-ph/9911261

\bibitem{lev wave error}
Levshakov, S.A., Tytler, D. \& Burles, S. 2000,
in {\em Early Universe: Cosmological Problems and Instrumental
Techniques}
proceedings of the Gamov Memorial Intern. Conf. St. Petersburg, Aug
23-28 1999, and Astr. \& Astrophysics Trans., vol. 19,
No. 3-4, pp. 385-396, Dec 2000,
astro-ph/9812114


\bibitem{linsky}
Linksy, J. 1998, Space Science Reviews, {\bf84}, 285

\bibitem{lopez_turner}
Lopez, R. \&\ Turner, M.S. 1999, Phys. Rev. D. {\bf59}, 103502

\bibitem{Lu_nitrogen}
Lu, L., Sargent, W.L.W. \& Barlow, T. 1998, AJ, {\bf115}, 55

\bibitem{d in gal chem}
Matteuci, F., Romano, F. \& Molaro, P. 1999 A\& Ap, {\bf 341},458

\bibitem[]{molarodh} Molaro, P., Bonifacio, P., Centurion, M., \&
Vladilo, G.  1999, preprint, astro-ph/9908060

\bibitem[]{nol00 errors}
 Nollett, K.M. \& Burles, S. Phys.Rev.D in press astro-ph/0001440
(2000).

\bibitem{oli_bbn_rev}
Olive, K.A., Steigman, G. \& Walker, T.P. 2000, Phys. Rept., {\bf333-334},
389

\bibitem{cbi cmb interferometer}
Padin, S., Cartwright, J.K., Mason, B.S., Pearson, T.J., Readhead, A.C.S.,
Shepherd, M.C., Sievers, J., Udomprasert, P.S., Holzapfel, S.W.L., Myers, S.T., 
Carlstrom, J.E., Leitch, E.M., Joy, M., Bronfman, L. \& May, J. 2001
ApJL submitted.

\bibitem{pagel_he_review}
Pagel, B.E.G. 2000, Phys. Rept., {\bf333-334}, 433

\bibitem{pra d in gal chem}
Prantzos, N. 1996 A\& Ap, {\bf 310}, 106

\bibitem{d in gal chem 2}
Prantzos, N. \& Silk, J. 1998 ApJ, {\bf 507},229

\bibitem{ryanli7 corrections}
Ryan, S., Beers, T.C., Olive, K.A., Fields, B.D. \& Norris, J.E. 2000
ApJ {\bf 530}, 57

\bibitem{ryanli7}
Ryan, S., Norris, J.E. \& Beers, T.C. 1999, ApJ {\bf 523}, 654 
astro-ph/9903059

\bibitem{savage}
Savage, B. \&\ Sembach, K.  1996 Annu. Rev. Astron. Astrophys. {\bf
34}, 279

\bibitem{sch98a}
Schramm, D. N. \& Turner, M. S., Rev. Mod. Phys {\bf70}, 303-318
(1998).

\bibitem{suz2000}
Suzuki, N. \& Tytler, D. 2000, in preparation

\bibitem{tosi}
Tosi, M., Steigman, G., Matteucci, F. \& Chiappini, C.,
Astrophys. J. {\bf498}, 226 (1998).

\bibitem[]{tytler96b} Tytler, D., Fan, X.M., \& Burles, S. 1996,
        Nature, {\bf 381}, 207

\bibitem{tytler97}
Tytler, D. \& Burles, S., 1997, in {\it Origin of Matter and
Evolution of Galaxies}, 
eds. T. Kajino, Y. Yoshii \& S. Kubono {\em (World Scientific Publ.
Co.)}, p.37-63.  astro-ph/9606110 


\bibitem[]{tytler99a} Tytler, D., Burles, S., Lu, L., Fan, X.M.,     
Wolfe, A., \& Savage, B.D. 1999, AJ, {\bf117}, 63


\bibitem[]{tytler2000} Tytler, D., O'Meara, J.M., Suzuki, N, \& Lubin,
D. 2000, 
{\em Physics Scripta}, {\bf 60}, in press, 
astro-ph/0001318 

\bibitem[Vogt etal 1994]{vog94}
Vogt, S. S. et al. 1994, Proc. SPIE, {\bf2198}, 362

\bibitem{walker bbn review 91}
Walker, T. P., Steigman, G., Schramm, D. N., Olive, K. A. \& Kang, H. S. 
1991 ApJ {\bf376}, 51


\bibitem{web97a}
Webb, J. K., Carswell, R. F., Lanzetta, K. M., Ferlet, R., Lemoine, M.,
Vidal-Madjar, A., \& Bowen, D. V. 1997, Nature, {\bf388}, 250

\bibitem{webb_VPFIT}
Webb, J. K. 1987, Ph.D. thesis, University of Cambridge

\bibitem{wolfe2000a}
Wolfe, A. \& Prochaska, J. X., preprint, astro-ph/009081


\end{thebibliography}
\end{document}